\documentclass{jfm}

\usepackage{graphicx}
\usepackage{newtxtext}
\usepackage{newtxmath}
\usepackage{natbib}
\usepackage{physics}
\usepackage{hyperref}
\hypersetup{
    colorlinks = true,
    urlcolor   = blue,
    citecolor  = black,
}

\newcommand{\RomanNumeralCaps}[1]
\linenumbers

\usepackage{orcidlink}

\def \TITLEnew {Theoretical framework for designing phase change material systems}

\def \Ra {\textit{Ra}}
\def \Pr {\textit{Pr}}
\def \St {\textit{St}}

\def \Pe {Pe}

\def \ST {{\textit{S}}_T}
\def \Nu {\textit{Nu}}
\def \Fo {\textit{Fo}}
\def \lp {\left(}
\def \rp {\right)}
\def \f {f}
\def \fc {\f^{\ast}}
\def \Rac {{\textit{Ra}}_e^{\ast}}

\def \ie {\textit{i.e.},~}

\def \ED {\textit{ED}}

\def \phiv {\Psi}

\title{\TITLEnew}

\author{Min Li~
\aff{1}
 \and Lailai Zhu~
 \aff{1}
  \corresp{\email{lailai\_zhu@nus.edu.sg}}}

\affiliation{\aff{1} Department of Mechanical Engineering,
National University of Singapore, 117575, Singapore}

\begin{document}
\maketitle

\begin{abstract}
Phase change materials (PCMs) hold considerable promise for thermal energy storage applications. However, designing a PCM system to meet specific performance  presents a formidable challenge, given the 
intricate influence of multiple factors on the performance.
To address this challenge, we hereby develop a theoretical framework that elucidates the melting process of PCMs. 
By integrating stability analysis with theoretical modeling,  we derive a transition criterion to demarcate different melting regimes, and subsequently formulate the melting curve that uniquely characterizes the performance of an exemplary PCM system. 
This theoretical melting curve captures the key trends observed in experimental and numerical data across a broad parameter space, establishing a convenient and quantitative relationship between design parameters and system performance.
Furthermore, we demonstrate the versatility of the theoretical framework across diverse configurations.
Overall, our findings deepen the understanding of thermo-hydrodynamics in melting PCMs, thereby facilitating the evaluation, design, and enhancement of PCM systems. 
\end{abstract}



\section{Introduction}
\label{sec:intro}

 Due to their high energy storage density, nearly constant melting temperature, and zero-carbon emissions, solid-liquid phase change materials (PCMs) have increasingly been utilized across various domains, including cold chain logistics \citep{tong2021phase, zhou2023adaptive}, thermal comfort in buildings \citep{kishore2023finned,ragoowansi2023realistic}, thermal energy storage \citep{gerkman2020toward,zhang2023accelerating}, and electronic thermal management \citep{wang2022fluidic,liu2022high}. The performance of  a PCM system is primarily governed by its energy storage capacity and power density \citep{gur2012searching,woods2021rate,yang2021phase,fu2022high,chen2022advanced},  which are influenced by various design parameters embedded in  material properties, system geometries, and operational conditions \citep{woods2021rate,yang2021phase}. Consequently, achieving the desired performance under specific conditions necessitates meticulous selection of storage materials and thoughtful design of system geometries. However, universal guidelines for this practice are not yet established 
~\citep{gao2021machine,yang2021phase,wang2022critical,he2022performance}, calling for deeper insights into the physical principles governing how design parameters impact system performance.

 The performance indicators---energy storage capacity  and power density---are determined by the maximum melted amount  and melting rate of PCMs, respectively. Thus, the  performance can be exclusively represented by  the temporal evolution of PCM's melted volume, which, in dimensionless form, is the  evolution of liquid fraction (melting curve). Indicatively, the task  of investigating the effects of design parameters on performance  can be restated as identifying their impacts on the melting curve.

 The melting curve has been typically recognised a composite of two segments, corresponding to successive regimes underwent by the storage process:
initially dominated by conduction and then by convection  \citep{ho1984heat,gau1986melting,wang1999experimental,duan2019melting,li2022melting}. 
Despite this understanding, a comprehensive 
model for the melting curve is lacking\citep{verma2008review,dutil2011review},  as current knowledge of the transitional scenario remains scarce~\citep{azad2021naturalI,azad2021natural,azad2022natural}, 
commonly limited to empirical fits from experimental or numerical data~\citep{wang1999experimental,vogel2016natural,duan2019melting,li2023melting}.

In this work, we delineate the transitional scenario of melting PCM and subsequently  develop a theoretical framework to model the melting curve. 
The resulting model encapsulates the essential physics of PCM melting, concomitantly agreeing  well with experimental and numerical data. The associated theoretical formula quantifies the impacts of  material properties, geometric features,  and operational conditions on system performance, providing convenient guiding principles for material selection and geometric design of PCM systems.

\section{Storage performance of PCM system}\label{Storage performance}

\subsection{Problem description} \label{physical model}

\begin{figure}
  \centerline{\includegraphics{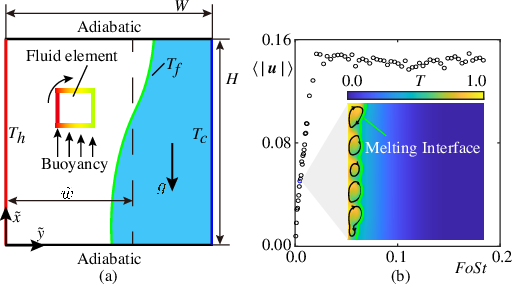}}
  \caption{(a) A rectangular cavity filled with subcooled PCM, maintaining temperatures at $T_h$ (hot) and $T_c$ (cold) at the left and right boundaries, respectively.
(b) Time evolution of the domain averaged velocity { $\langle |\boldsymbol{ u}| \rangle=(g\alpha \Delta T_l H)^{-1/2}\int |\boldsymbol{\tilde u}| {\rm d} {\mathcal{V} } /\mathcal{V}_0$,}  where $\mathcal{V}_0=WH$. Here, ${\Ra}=10^7$, ${\Pr}=0.1$, ${\St}=0.1$, $\ST=1.0$, and $\gamma=1.0$. 
The inset shows the instantaneous temperature field and streamlines at ${\Fo}{\St}=0.0046$.  }
\label{fig:Physical model}
\end{figure}

We illustrate the theoretical framework using a minimal yet representative enclosure of PCM~\citep{dhaidan2015melting}---a rectangular cavity  of width $W$ and height $H$ (see figure~\ref{fig:Physical model}a). It is filled with solid PCM initialized at the ambient temperature  $\tilde T=T_c$, which is below the fusion (melting) temperature $T_f$.  Keeping its top and bottom walls insulated, the cavity's left wall absorbs heat from a hot working device at temperature  $\tilde T=T_h>T_f$, with its right boundary subject to a cold ambient environment at $\tilde T=T_c<T_f$. 
This configuration finds typical application in enhancing thermal comfort in buildings \citep{lachheb2024enhancing}, where PCM is encapsulated within bricks, with  $T_f$ lying between indoor ($T_c$) and outdoor ($T_h$) temperatures.   The ideal configuration---where $T_f$ always equals $T_c$, preventing any heat transfer in the solid phase---has been discussed in previous studies \citep{esfahani2018basal,favier2019rayleigh,yang2022abrupt,li2022melting}.

Using tildes to indicate dimensional variables,
the thermo-hydrodynamic equations for the liquid phase read~
\begin{subequations}
\begin{align}
\grad \cdot \boldsymbol {\tilde u} & =0, 
\label{mass conservation}\\
\frac{\partial{\boldsymbol{ \tilde{u}} }}{\partial \tilde t}+\boldsymbol{{ \tilde{u}} }\cdot\grad\boldsymbol { \tilde{u}} & =\boldsymbol{g}\alpha({ \tilde{T}} -T_0)+ {\nu \nabla^2 \boldsymbol{{ \tilde{u}} }}-\frac{\grad{{ \tilde{p}} }}{\rho_0}, 
\label{momentum conservation} \\
\frac{\partial{\tilde T}}{\partial  \tilde t}+\boldsymbol{\tilde u}\cdot\grad \tilde{T} & = {\kappa \nabla^2 \tilde{T}},
\label{energy conservation}
\end{align}
\label{eq:ns_T}
\end{subequations}
where  $\boldsymbol{\tilde u}$, $\tilde T$, and $\tilde p$ denote the velocity, temperature, and pressure of the liquid, respectively; its kinematic viscosity is $\nu$, while the thermal expansion coefficient and diffusivity are  $\alpha$ and $\kappa$, respectively. Here, $\rho_0$ denotes the reference density measured at the reference temperature $T_0=(T_h+T_f)/2$, and  $\boldsymbol{g}$ indicates the gravitational acceleration. At the melting interface between the solid ($s$) and liquid ($l$) phase, the Stefan condition $\mathcal{L} \boldsymbol{\tilde V}/C_p=[(\kappa\grad \tilde T)_s-(\kappa\grad \tilde T)_l]\cdot \boldsymbol{n}$ holds, where $\mathcal{L}$ and $C_p$ are the specific latent heat and specific heat, respectively;   $\boldsymbol{\tilde V}$ represents the velocity of the interface, and $\boldsymbol{n}$ denotes its liquid-facing unit normal.  

The above governing equations denote that the material properties affecting the system performance include $\alpha$, $T_f$, $\nu$, $\kappa$, $C_p$, and $\mathcal{L}$. Their effects, along with those of the geometry ($W$ and $H$), and the working conditions ($T_h$ and $T_c$) are characterised by five dimensionless numbers: Rayleigh number ${\Ra}$, Prandtl number $\Pr$, Stefan number ${\St}$, subcooling strength $\ST$, 
and the cavity's aspect ratio $\gamma$, defined as
\begin{equation}
{\Ra}=\frac{g\alpha \Delta T_l H^3}{\nu\kappa},~ {\Pr}=\frac{\nu}{\kappa},~\St=\frac{ C_p \Delta T_l}{\mathcal{L}},
\ST=\frac{\Delta T_s}{\Delta T_l},~\gamma=\frac{W}{H}.
\label{control parameters}
\end{equation}
Here,  $\Delta T_s=T_f-T_c$ and $\Delta T_l=T_h-T_f$ are the temperature differences across the solid and liquid phases, respectively.
Additionally, the melting behavior of PCMs is characterized by the Nusselt number ${\Nu}$, Fourier time ${\Fo}$, and liquid fraction $\f$,  
\begin{equation}
{\Nu}=-\frac{H}{\Delta T_l}\langle \frac{\partial \tilde T}{\partial \tilde y} \rangle_{\tilde y=0},~ {\Fo}=\frac{\kappa \tilde t}{W^2},~\f=\frac{\tilde w}{W},
\label{response parameters}
\end{equation}
where $\tilde w$  is the average thickness of the liquid layer (figure~\ref{fig:Physical model}a),  and $\langle~ \rangle_{\tilde y=0}$ denotes spatial averaging along the hot wall at $\tilde y = 0$.

\subsection{Linear stability analysis} \label{lsa}

We obtain the transition criterion through linear stability analysis, choosing the endpoint of conduction-dominated regime as the base state. 
Since the horizontal  temperature gradient imposes a heterogeneous buoyancy on the fluid (figure~\ref{fig:Physical model}a), the  corresponding base flow $\boldsymbol{\tilde{u}}_b $ is not stationary, as  evidenced by the temporal evolution of the domain-averaged velocity $\langle |\boldsymbol{u}| \rangle$ in figure~\ref{fig:Physical model}b. This 
non-stationary
base flow $\boldsymbol{\tilde{u}}_b$ differs from the quiescent counterpart of the canonical Rayleigh-B\'enard convection, and cannot be derived analytically. However,  the base liquid temperature $\tilde{T}_b$ is readily calculable, because the convective heat transfer is negligible in the conduction-dominated regime.
By solving the 
heat conduction
equation, we obtain $\tilde{T}_b = T_h-{\Delta T_l} \tilde y /{\tilde w_b}$, 
where $\tilde w_b = W/(1 + \ST^{\ast}) $  results from the energy balance between the liquid and solid phases, 
$\Delta T_l/\tilde w_b= \Delta T_s/\left( W-\tilde w_b \right) $. 
Here, the superscript $^{\ast}$ indicates that $\ST^\ast$ is the critical subcooling strength required for the base state to reach the onset of instability.

To linearise \eqref{eq:ns_T}, we decompose the flow and temperature fields into the base state $\lp \tilde{\boldsymbol{u}}_b, \tilde{p}_b, \tilde T_b\rp$ and the disturbance state $\lp \tilde{\boldsymbol{u}}^{\prime}, \tilde{p}^{\prime}, \tilde T^{\prime}\rp $, such that $\lp \tilde{\boldsymbol{u}}, \tilde{p}, \tilde T\rp = \lp \tilde{\boldsymbol{u}}_b, \tilde{p}_b, \tilde T_b\rp+\lp \tilde{\boldsymbol{u}}^{\prime}, \tilde{p}^{\prime}, \tilde T^{\prime}\rp$. Selecting $\tilde w_b$  as the length scale,  $U_0$ and $U$ the base and perturbation velocity scales, respectively, we derive the dimensionless linearised equations for $\lp \tilde{\boldsymbol{u}}^{\prime}, \tilde{p}^{\prime}, \tilde T^{\prime}\rp $,
\begin{subequations}
\begin{align}
\grad \cdot \boldsymbol u'  &=0  , 
\label{dimensionless perturbation mass conservation}\\
\frac{\partial{\boldsymbol{ u}' }}{\partial t} +  \frac{U_0 \tilde w_b}{\kappa} (\boldsymbol{ u }_b \cdot\grad{\boldsymbol{u}'}+\boldsymbol{ u }' \cdot\grad{ \boldsymbol{u}_b} )  &= \frac{ {\Ra}{\Pr}\gamma^3}{(1+\ST^{\ast})^3} T' \boldsymbol{e_x}+ {\Pr}  {\nabla^2} \boldsymbol{ u }' -\grad{{ p'} },
\label{dimensionless perturbation momentum conservation}\\
\frac{\partial{T'}}{\partial t} -  v' + \frac{U_0 \tilde w_b}{\kappa}\boldsymbol{u}_b\cdot \grad T'  &=  {\nabla^2} T',
\label{dimensionless perturbation energy conservation}
\end{align}
\label{perturbation equations}
\end{subequations}
where we have also chosen  $\tilde w_b^2/\kappa$ as the characteristic time, $\rho_0 \kappa U/\tilde w_b$ the characteristic pressure, and $\Delta T_l U \tilde w_b /\kappa$ the characteristic temperature. Here, $v^{\prime}$ denotes the $y$-component of the dimensionless perturbation velocity $\boldsymbol{u}^{\prime}$.

We note that $U_0\tilde w_b/\kappa$ in ~\eqref{dimensionless perturbation momentum conservation} and \eqref{dimensionless perturbation energy conservation} represents the P\'eclet number, which measures the relative strength of convection to conduction. 
{Since this number is inherently small in the conduction-dominated base state, we simplify the theoretical analysis by neglecting all terms involving $U_0\tilde w_b/\kappa$ in ~\eqref{perturbation equations}. This assumption proves reasonable, as will be verified \textit{a posterior} by numerical simulations (see appendix \ref{Appendix Small Pelect number}). 
With this simplification, we obtain}
\begin{subequations}
\begin{align}
\grad \cdot \boldsymbol u'  &=0  , 
\label{dimensionless perturbation mass conservation2}\\
\frac{\partial{\boldsymbol{ u}' }}{\partial t}  &= \frac{ {\Ra}{\Pr}\gamma^3}{(1+\ST^{\ast})^3} T' \boldsymbol{e_x}+ {\Pr}\nabla^2 \boldsymbol{ u }' -\grad{{ p'} },
\label{dimensionless perturbation momentum conservation2}\\
\frac{\partial{T'}}{\partial t} -  v'   &=  \nabla^2 T'.
\label{dimensionless perturbation energy conservation2}
\end{align}
\label{perturbation equations2}
\end{subequations}

Using the normal mode approach,   a dimensionless perturbation field $\phi$ of either velocity, temperature, or pressure is expressed as $\phi' (x,y,t)=\hat \phi(y)\exp({\rm i}\mathcal{K}x+\sigma t)$, where $\hat \phi$ presents the complex magnitude of perturbation,  $\mathcal{K}$ and $\sigma$  stand for the wavenumber and growth rate, respectively; here, ${\rm i}=\sqrt{-1}$. Substituting the normal modes into ~\eqref{perturbation equations2}, we obtain the governing equations for the perturbation magnitudes,  
\begin{subequations}
\begin{align}
0&={\rm i}\mathcal{K}\hat u +\frac{\partial \hat v}{\partial y},
\label{mass conservation for perturbation magnitude}\\
\sigma \hat u &=\frac{{\Ra}{\Pr}\gamma^3}{(1+{\ST^{\ast}})^3} \hat T +{\Pr}\left(\frac{\partial^2 \hat u}{\partial y^2} -\mathcal{K}^2 \hat u\right)-{\rm i}\mathcal{K}\hat p, 
\label{x momentum conservation for perturbation magnitude}\\
\sigma \hat v &= {\Pr} \left(\frac{\partial^2 \hat v}{\partial y^2}-\mathcal{K}^2\hat v \right)-\frac{\partial \hat p}{\partial y},
\label{y momentum conservation  for perturbation magnitude}\\
\sigma \hat T &=  \left( \frac{\partial^2 \hat T}{\partial y^2}-\mathcal{K}^2\hat T \right)+\hat v.
\label{energy conservation  for perturbation magnitude}
\end{align}
\label{perturbation magnitude equations}
\end{subequations}
The corresponding boundary conditions at the hot wall $y=0$ are
\begin{align}
\hat u = \hat v =0,~ \hat T =0.
\label{BCs at hot wall}
\end{align}
Those at the melting interface $y=1$ read
\begin{align}
\hat u = \hat v =0,~  \mathcal{K} \hat T {\rm coth}(\mathcal{K} {\ST^{\ast}}) + \frac{\partial \hat T}{\partial y}=0,
\label{BCs at the melting interface}
\end{align}
where the condition for $\hat{T}$ is derived from the Stefan condition \citep{davis1984pattern,toppaladoddi2019combined,lu2022rayleigh}.

Equations~\eqref{perturbation magnitude equations}, \eqref{BCs at hot wall}, and \eqref{BCs at the melting interface} indicate that the transition depends on ${\Pr}$, ${\Ra}$, $\gamma$, and ${\ST^{\ast}}$. 
To investigate the influence of these parameters, 
we simplify ~\eqref{perturbation equations2} into a single equation representing the marginal state ($\sigma=0$).
Taking the divergence of ~\eqref{dimensionless perturbation momentum conservation2} and 
{using}~\eqref{dimensionless perturbation mass conservation2}, we derive
\begin{equation}
0 =\frac{{\Ra}{\Pr}\gamma^3}{(1+\ST^{\ast})^3} \frac{\partial T'}{\partial x}-{\nabla^2}{{ p'} }.
\label{dimensionless perturbation momentum conservation with divergence}
\end{equation}
Next, differentiating ~\eqref{dimensionless perturbation momentum conservation with divergence} with respect to $y$ results in
\begin{equation}
0 =\frac{{\Ra}{\Pr}\gamma^3}{(1+\ST^{\ast})^3} \frac{\partial^2 T'}{\partial x \partial y}-{\nabla^2}{\frac{\partial p'}{\partial y} }.
\label{dimensionless perturbation momentum conservation with divergence and differentiation}
\end{equation}
We then take the Laplacian of the $y$-component of  ~\eqref{dimensionless perturbation momentum conservation2} and obtain
\begin{equation}
\frac{\partial}{\partial t} {\nabla^2} {v' }={\Pr}{\nabla^4} v'-{\nabla^2}\frac{\partial p'}{\partial y}.
\label{dimensionless perturbation momentum conservation for y component}
\end{equation}
Further subtracting ~\eqref{dimensionless perturbation momentum conservation with divergence and differentiation} from ~\eqref{dimensionless perturbation momentum conservation for y component} yields:
\begin{equation}
\frac{\partial}{\partial t} {\nabla^2} {v' }={\Pr}{\nabla^4} v'-\frac{{\Ra}{\Pr}\gamma^3}{(1+\ST^{\ast})^3}  \frac{\partial^2 T'}{\partial x \partial y}.
\label{dimensionless perturbation momentum conservation without p}
\end{equation}

Incorporating the normal modes into ~\eqref{dimensionless perturbation energy conservation2} and ~\eqref{dimensionless perturbation momentum conservation without p}, the perturbation magnitudes  at the marginal state ($\sigma=0$) are governed by
\begin{subequations}
\begin{align}
\mathcal{K}^2\hat T-\hat v-\hat T^{(2)}=0,
\label{dimensionless perturbation of T}\\
-{\rm i}\mathcal{K}\gamma^3{\Ra}\hat T^{(1)}+\mathcal{K}^2(1+\ST^\ast)^3(\mathcal{K}^2\hat v-2\hat v^{(2)})+(1+\ST^\ast)^3\hat v^{(4)}=0,
\label{dimensionless perturbation of v}
\end{align}
\label{two perturbation magnitude}
\end{subequations}
{where $\hat T^{(n)} = \frac{\partial^n \hat T}{\partial y^n}$ and $\hat v^{(n)} = \frac{\partial^n \hat v}{\partial y^n}$.
}
By eliminating $\hat v$,  we derive
\begin{equation}
\mathcal{K}^6\hat T - \frac{{\rm i}{\Ra}\gamma^3 \mathcal{K}\hat T^{(1)}}{(1+{\ST^\ast})^3} =3\mathcal{K}^4\hat T^{(2)}-3\mathcal{K}^2\hat T^{(4)}+\hat T^{(6)},
\label{Single perturbation equation}
\end{equation}
which suggests that the transition is  characterized solely   by a critical 
effective Rayleigh number  $\Rac={\Ra} \gamma^3/(1+{\ST^\ast})^3$.

\subsection{Results and discussions} \label{result}

\begin{figure}
  \centerline{\includegraphics{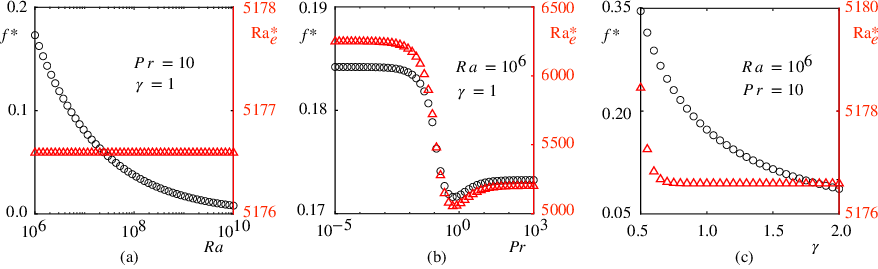}}
  \caption{Typical dependence of the critical liquid fraction $\fc$ and critical effective Rayleigh number $\Rac$ on (a) $\Ra$ with ${\Pr}=10$ and $\gamma=1$, (b) $\Pr$ with $\Ra=10^6$ and $\gamma=1$, and (c) $\gamma$ with  $\Ra=10^6$ and ${\Pr}=10$. }
\label{fig:Stability curve}
\end{figure}

 Realizing the critical liquid fraction $\fc=1/\lp 1+\ST^{\ast}  \rp $
upon substituting $\tilde w_b=W/(1+\ST^{\ast})$ into ~\eqref{response parameters}, we further obtain
 \begin{equation}
\fc=\left(\frac{\Rac}{{\Ra} \gamma^3}\right)^{1/3}.
\label{transition criterion}
\end{equation}
{Applying  a spectral method~\citep{weideman2000matlab} to \eqref{perturbation magnitude equations}--\eqref{BCs at the melting interface}, we calculate $\fc=1/\lp 1+\ST^{\ast}  \rp $  for various configurations within the ranges of $10^6 \leq \Ra \leq 10^{10}$, $10^{-5} \leq \Pr \leq 10^3$, and $0.5 \leq \gamma \leq 2$, further yielding $\Rac$ via \eqref{transition criterion}.} The results consistently exhibit a similar trend across these ranges. This trend is revealed by a representative set of data presented in figure~\ref{fig:Stability curve}.

 Figure~\ref{fig:Stability curve}a shows that $\Rac$ remains constant with respect to ${\Ra}$. Additionally, $\Rac$ is independent of $\Pr$ when ${\Pr}<10^{-3}$ and ${\Pr}> 10$, 
plateauing at $6248$ and $5200$, respectively (figure~\ref{fig:Stability curve}b). 
Notably, $\fc$ corresponding to both the highest and lowest values  of $\Rac$  are closely matched,  as indicated by $\lp 6248/5200\rp^{1/3}\approx 1.06$. Consequently, in  the subsequent theoretical modelling, $\fc$  can be  approximated using the  median value of $\Rac=5724$.
However, such an approximation fails for narrow (small $\gamma$) cavities, where $\Rac$ increases rapidly when $\gamma$ decreases from $0.75$, as evidenced in figure~\ref{fig:Stability curve}c. 
This failure stems from the stronger confinement effect, as commonly observed in low-aspect-ratio canonical Rayleigh-B\'ernard convection~\citep{huang2013confinement,huang2016effects,yu2017onset,wang2012linear,shishkina2021rayleigh,ahlers2022aspect}.

Having derived the transition criterion $\fc$, we employ it to develop the mathematical model for the melting curve.  Following \citet{esfahani2018basal,favier2019rayleigh,yang2022abrupt}, we start from the dimensional energy equation for both phases {  \citep{voller1987enthalpy, huang2013new}},
\begin{equation}
\frac{\partial{\tilde T}}{\partial  \tilde t}+\boldsymbol{\tilde u} \cdot \grad    \tilde{T} ={\kappa \nabla^2 \tilde{T}} -\frac{\mathcal{L}}{C_p}\frac{\partial {\f_{loc}}}{\partial \tilde t},
\label{energy equation for the whole cavity}
\end{equation}
{where $f_{loc}$ represents the local liquid fraction [see \eqref{local fl}] and its volume average is equivalent to the macroscopic counterpart $f$ in \eqref{response parameters}}. 
The first and last terms represent the change rates of the sensible and latent heats, respectively. 
The ratio of the two terms
, \ie
the Stefan number, 
is typically far below unity; namely,
$\St=C_p\Delta T_l/\mathcal L\ll 1$, because of the high $\mathcal L$ of PCMs. This 
condition as commonly satisfied in prior  studies~\citep{beckermann1989effect,ho2013experimental,arasu2012numerical,kean2019numerical,regin2009analysis,khan2019thermodynamic}, 
allowing for neglecting the first term.

Further applying Gauss's theorem, we obtain the integral form of ~\eqref{energy equation for the whole cavity},
\begin{equation}
\frac{\mathcal{L}}{C_p}\frac{\partial \f}{\partial \tilde t} W =\kappa \left[\langle\frac{\partial \tilde T}{\partial \tilde y}\rangle_{\tilde y=W} - \langle\frac{\partial \tilde T}{\partial \tilde y}\rangle_{\tilde y=0} \right],
\label{fl equation}
\end{equation}
where $\langle\partial \tilde T/ \partial \tilde y \rangle_{\tilde y=W}$, as the average temperature gradient at the cold wall, varies over time generally due to the advancing melting interface. 
However, in the limit of ${\St}\ll 1$ as revealed above, the interfacial velocity is negligible compared to the conduction rate, as implied by the normalized Stefan condition $\boldsymbol{\tilde V}/({\kappa}/H)={\St}~ (\grad  T_{s}+ \grad  T_{l})\cdot \boldsymbol{n}$.
{
Consequently, the conduction process can be considered quasi-steady on the timescale of the interfacial evolution, leading the temperature gradient at the cold wall to 
}
\begin{equation}
\langle\frac{\partial \tilde T}{\partial \tilde y}\rangle_{\tilde y=W}=-\frac{ \Delta T_s}{W-\tilde {w}},
\label{T gradient at cold wall}
\end{equation}
where $W-\tilde w$ quantifies the average thickness of the solid phase (see figure~\ref{fig:Physical model}a). 

 At the hot wall, the temperature gradient $\langle \partial \tilde T/ \partial \tilde y \rangle_{\tilde y=0}$  depends on the conduction and convection states of the liquid phase. When conduction dominates, this gradient becomes $-\langle \partial \tilde T/\partial \tilde y\rangle_{\tilde y=0}=\Delta T_l/\tilde w$, which  combining ~\eqref{fl equation} and \eqref{T gradient at cold wall}  yields the conduction melting curve in the implicit form
\begin{equation}
 \frac{\f(1+\ST)(\f-2\ST+\f\ST)-2\ST\ln[1-\f(1+\ST)]}{2(1+\ST)^3} 
= \Fo\St.
\label{conduction fl}
\end{equation}
Due to the  subcooling effects, this solution differs from the classical Stefan solution $\f \sim \sqrt{\Fo}$~\citep{esfahani2018basal,fu2022high}, but recovers to it when $\ST=0$.

\begin{figure}
  \centerline{\includegraphics{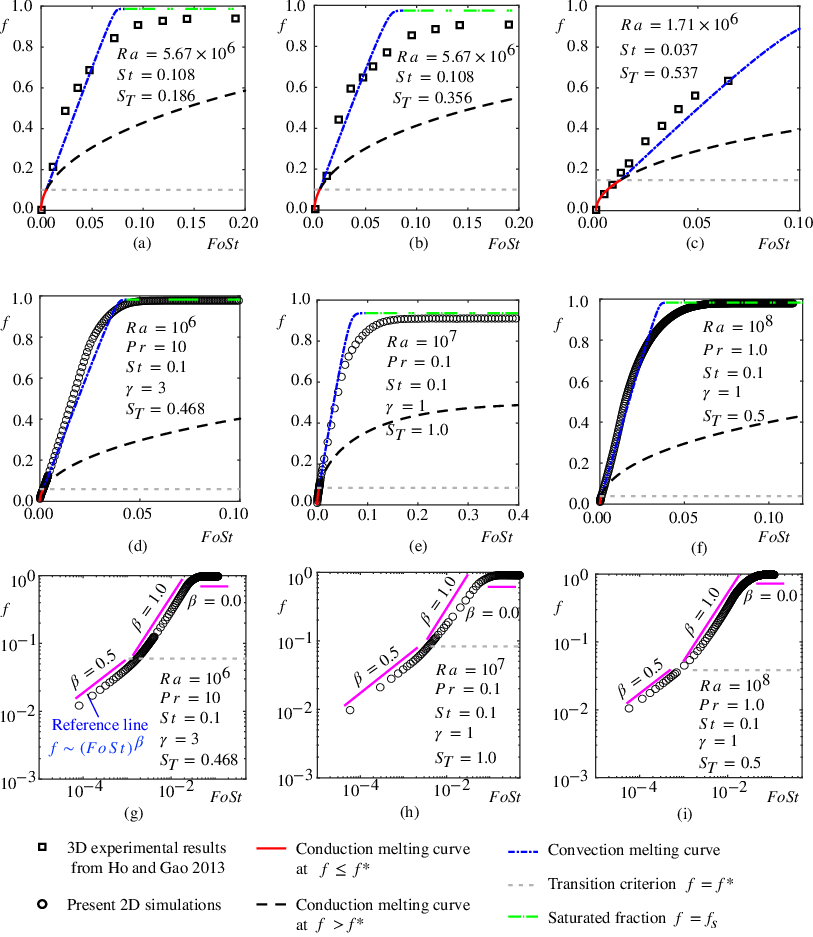}}
  \caption{The present model is compared with (a--c)  experimental measurements (squares) of n-octadecane ($\Pr=56$) in a cubic cavity~\citep{ho2013experimental}, and (d--f) our two-dimensional  numerical data (circles) for a rectangular cavity. {The numerical data in (d--f) is represented on a log-log scale in (g--i) to highlight the three melting regimes.}
  }
\label{fig:validation}
\end{figure}

As convection dominates, the temperature gradient can be modelled as $-\langle \partial \tilde T/\partial \tilde y \rangle_{\tilde y=0}={\Nu} \Delta T_l/H$ following~\eqref{response parameters}, resulting in the convection melting curve:
\begin{equation}
\frac{\f}{{\Nu}\gamma}-\frac{\ST}{({\Nu}\gamma)^2 }\ln \left[ {\Nu}\gamma(\f-1)+\ST \right]= \Fo \St +C,
\label{convection fl}
\end{equation}
where $C$ is a constant to be determined. 
Because   both melting  curves concatenate at $\left( {\Fo}^{\ast} \St,\fc \right)$,
subtracting ~\eqref{conduction fl} from ~\eqref{convection fl} yields the constant
\begin{eqnarray}
C &=&\frac{\fc}{{\Nu}\gamma}-\frac{\ST}{({\Nu}\gamma)^2 }\ln \left[ {\Nu}\gamma(\fc-1)+\ST \right] \nonumber\\
&-&\frac{\fc(1+\ST)(\fc-2\ST+\fc\ST)-2\ST\ln[1-\fc(1+\ST)]}{2(1+\ST)^3},
\label{dimensionless constant C}
\end{eqnarray}
where the Nusselt number is expressed using established models, specifically 
$\Nu \approx 0.5{\Ra}^{1/4}{\Pr}^{1/4}$ for ${\Pr}\leq 0.1$ and $\Nu \approx 0.24{\Ra}^{0.26}{\Pr}^{0}$ for $ \Pr> 0.1$~\citep{beckermann1989effect,shishkina2016momentum,wang2021regime}.

 By  substituting ~\eqref{dimensionless constant C} into ~\eqref{convection fl},  the domain of the logarithmic function necessitates 
\begin{equation}
\f < \f_s=1-\frac{\ST}{{\Nu} \gamma}.
\label{definition domain}
\end{equation}
This indicates that  melting will cease at a saturated liquid fraction $\f_s$, leaving  a portion of solid PCM  unmelted. 

Thus far, we have derived the critical liquid fraction $\fc$ and established a mathematical model for the melting curve.
As shown in figure~\ref{fig:validation}a--c, the theoretical melting curves agree  well with the prior experimental data~\citep{ho2013experimental}. Besides, these curves also capture the trend revealed in our two-dimensional simulations, see figure~\ref{fig:validation}d--f. In the long-time limit, these curves asymptotically flattens, converging to $f_s$ as expected.
{To better characterize the transition, we depict these curves on a log-log scale in figure~\ref{fig:validation}g--i. The temporal evolution of $f$ exhibits three distinct phases: the first two follow a power-law relationship $\f \sim (\Fo \St)^\beta$ with exponents $\beta\approx 0.5$ and $\beta \approx 1.0$, respectively, followed by a saturated phase. 
Notably, this power-law scaling is exact in no-subcooling cases ($\ST=0$), where $\beta=0.5$ and $1.0$ correspond to the conduction-dominated and convection-dominated regimes, respectively~\citep{esfahani2018basal,favier2019rayleigh,li2022melting}. For present subcooling scenarios ($\ST>0$), the first
two regimes with different $\beta$ are precisely demarcated by the predicted transition criterion $f=\fc$ (gray dashed line).}
Our theory's predictive ability is further validated through benchmark comparisons with both our simulations and published data~\citep{beckermann1989effect} as detailed in Appendix \ref{Numerical method}.

Despite the generally satisfactory agreement between theoretical predictions and the experimental/numerical data, slight discrepancies are observed. They arise from two approximations: 1) neglecting the time variation of sensible heat  in \eqref{energy equation for the whole cavity} and 2) assuming a steady temperature gradient at the cold wall to obtain \eqref{T gradient at cold wall}. 
Both approximations hold when  $\St \ll 1$.  
At $\St = 0.1$,  figure  \ref{fig:validation} indicates that the maximum relative error between the theoretical and numerical predictions of $\f$ is $\approx 15\%$. This error increases with  $\St$, reaching $\approx 20\%$ when  $\St \in \lp 0.2, 0.4 \rp $, see Appendix \ref{Appendix Stefan's effect}.

We now exploit the theoretical melting curve $f(\Fo)$ to predict the key performance indicators. We start with the energy density $\ED$ as the cumulative energy stored per unit volume of PCM
\citep{woods2021rate,fu2022high}, 
which depends on the liquid fraction $f$ through
\begin{equation}
\ED\lp \tilde{t} \rp=\rho [C_p(T_f -T_c)+\mathcal{L}] \f .
\label{PCMs indicators}
\end{equation}
Here, we have neglected the liquid's sensible heat that plays a negligible role, as demonstrated in \citet{fu2022high} and confirmed by our experimentally-verified calculations (see Appendix \ref{inclined cavity}). Notably, the maximum value of  $\ED$, reached at $\f=f_s$, represents the energy storage capacity, whereas the temporal derivative of $\ED$ serves as another indicator, the power density \citep{woods2021rate,fu2022high}. 

Practically, ~\eqref {PCMs indicators}  
characterizes how design parameters of a PCM system affect its performance
via the dimensionless numbers encoded in the melting curve. 
This characterization, requiring only simple algebraic calculations, enables the rapid determination of parameters customized to meet specified performance criteria, thereby streamlining the system's design.

Besides precisely characterizing performance, the theoretical solutions also provide general guidance for optimizing design parameters. 
For example, the critical liquid fraction $\fc$ signifies the duration of the conduction-dominated regime, which features inefficient heat transfer. 
Hence, reducing  $\fc$  can enhance the melting rate \citep{sun2016experimental,azad2021naturalI}, \ie the power  density. 
In the current configuration, this can be achieved by increasing the aspect ratio $\gamma$, see   \eqref{transition criterion}.

\section{Conclusion} \label{conclusion}

By employing linear stability analysis and the integral energy equation, we have developed a theoretical framework for modeling the melting process and performance of PCM systems. Focusing on a exemplary system characterized by lateral heating, we derive an experimentally and numerically validated mathematical model for the melting curve. This curve facilitates rapid designs of PCM systems
tailored to specific performance targets. 
Furthermore, we illustrate, in Appendix \ref{NEPCM melting}--\ref{Annular Tube}, the theoretical framework's versatility by modelling a variety of other representative PCM configurations, with theoretical predictions corroborated by prior studies \citep{kean2019numerical,korti2020experimental,dhaidan2013experimental}. These configurations vary in material species---ranging from pure PCM to nanoparticle-enhanced PCM, in geometries---including an inclined rectangular cavity and an annular tube, and in operational conditions, such as isothermal and constant power heating. Overall, our investigation exemplifies the application of thermo-hydrodynamic theory and physics in facilitating the development of engineering solutions for energy applications.

\backsection[Funding]{L.Z. acknowledges the partial support received from the National University of Singapore, under the startup grant A-0009063-00-00. The computation of the work was performed on resources of the National Supercomputing Centre, Singapore (https://www.nscc.sg)}

\backsection[Declaration of interests]{
 The authors report no conflict of interest.}

\backsection[Author ORCID]{
Min Li, https://orcid.org/0000-0001-5785-6617; Lailai Zhu, https://orcid.org/0000-0002-3443-0709}

\appendix

\section{Numerical Method}\label{Numerical method}

In this section, we first present the numerical method used to simulate the melting process of PCM. We then validate the numerical solver using published data.

The numerical simulations of the melting process are performed using the lattice Boltzmann method \citep{huang2013new,luo2015lattice}, which employs density distribution  $\eta_i$ and temperature distribution $\zeta_i$  to describe the evolution of velocity and temperature, respectively. The evolution equations at grid location $\boldsymbol{x}$  are given by:
\begin{subequations}
\begin{align}
\eta_i(\boldsymbol{x}+\boldsymbol{e}_i, t+1) & =\eta_i(\boldsymbol{x}, t) + \frac{1}{\tau_v}\left[\eta_i^{eq}(\boldsymbol{x}, t) -\eta_i(\boldsymbol{x}, t)\right] +  F_i, 
\label{density distribution}\\
\zeta_i(\boldsymbol{x}+\boldsymbol{e}_i, t+1) & =\zeta_i(\boldsymbol{x}, t) + \frac{1}{\tau_T}\left[\zeta_i^{eq}(\boldsymbol{x}, t)-\zeta_i(\boldsymbol{x}, t) \right] + S, 
\label{tmepreature distribution} 
\end{align}
\label{eq:LBM}
\end{subequations}
where $\boldsymbol{e}_i$ is the discrete velocity in direction $i$, and  $\tau$  with different subscripts represents the dimensionless relaxation time.  
Here,  $F_i$ is the discretized body force and  $S$ represents the latent heat source.
The equilibrium distributions are expressed as
\begin{subequations}
\begin{align}
\eta_i^{eq}& =\tilde \rho \omega_i[ 1+3\boldsymbol{e}_i \cdot \boldsymbol{\tilde u}+4.5(\boldsymbol{e}_i \cdot \boldsymbol{\tilde u})^2 -1.5 \boldsymbol{\tilde u}^2], 
\label{feq}\\
\zeta_i^{eq}& =\tilde T \omega_i[ 1+3\boldsymbol{e}_i \cdot \boldsymbol{\tilde u}+4.5(\boldsymbol{e}_i \cdot \boldsymbol{\tilde u})^2 -1.5 \boldsymbol{\tilde u}^2]. 
\label{geq}
\end{align}
\label{eq:equilibrium}
\end{subequations}
The weight coefficient in direction $i$ is represented by $w_i$ . The density, velocity and temperature are calculated as
\begin{equation}
\tilde \rho =\sum_i \eta_i, ~\tilde \rho\boldsymbol{\tilde u}=\sum_i \eta_i\boldsymbol{e}_i,~\tilde T=\sum_i \zeta_i.
\label{Macro quantities}
\end{equation}
Notably, 
all quantities are expressed in lattice units, which can be converted to physical units according to specific rules \citep{feng2007coupled,xia2024particle}.

The discretized body force in ~\eqref{density distribution} reads  \citep{guo2013lattice},
\begin{equation}
F_i = 3 \omega_i \boldsymbol{e}_i \cdot [- \tilde \rho \boldsymbol{g}\alpha(\tilde T-T_0)].
\label{Body force}
\end{equation}
The source term in ~\eqref{tmepreature distribution}  is 
\begin{equation}
S = -\omega_i \frac{\mathcal L}{C_p} [{f_{loc}}(\boldsymbol{x},t)- {f_{loc}}(\boldsymbol{x},t-\Delta t)],
\label{phase-change term}
\end{equation}
which quantifies the propagation effects of melting interface. The local liquid fraction ${f_{loc}} (\boldsymbol{x},t)$ is  determined from the enthalpy interpolation
\begin{equation}
{f_{loc}} (\boldsymbol{x},t)=\left\{
             \begin{array}{lr}
             0, & ~{\rm if}~\mathcal H< \mathcal H_s=C_p T_f  \\
             \frac{\mathcal H- \mathcal H_s}{\mathcal H_l- \mathcal H_s},& ~{\rm if}~\mathcal H_s \leq \mathcal H \leq \mathcal H_l=\mathcal H_s+\mathcal L \\
             1, &  ~{\rm if}~\mathcal H>H_l,
             \end{array}
\right.
\label{local fl}
\end{equation}
where the enthalpy is calculated as $\mathcal H=C_p \tilde T +{f_{loc}}(\boldsymbol{x}, t-\Delta t)\mathcal L$.

By Chapman-Enskog analysis, the mesoscale evolution equations,   \eqref{density distribution} and \eqref{tmepreature distribution}, recover the macroscopic conservation laws in the limit of low Mach number  \citep{guo2013lattice,huang2013new}. Additionally, this analysis shows that  $\tau_v = 3\nu+0.5$ and $\tau_T=3\kappa+0.5$. 

\begin{figure}
  \centerline{\includegraphics{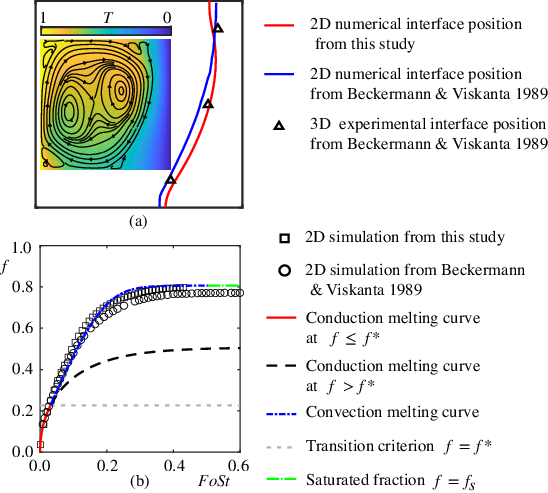}}
  \caption{Comparison between our numerical results and those from 
\citet{beckermann1989effect} when ${\Ra}=4.877\times 10^5$, ${\Pr}=0.0208$, ${\St}=7.557\times 10^{-2}$, $\gamma=1$, and $\ST=0.971$. (a) Interface position at the end of  the melting process, with the inset showing the temperature field and streamlines; (b) liquid fraction $f$ versus $\Fo\St$.}
\label{fig:codevalidation}
\end{figure}

Using the above method, we conduct two-dimensional direct numerical simulations for a lateral heating cavity.
The non-equilibrium extrapolation scheme~\citep{Guo2002non} is implemented to impose both the  velocity and temperature boundary conditions.
{Following previous studies \citep{favier2019rayleigh,purseed2020bistability},  we assume identical physical properties for the solid and liquid phases. }
\
Additionally, micro-scale Gibbs-Thomson effects are neglected, ensuring $\tilde T_l = \tilde T_s =T_f$ at the melting interface \citep{davis1984pattern, toppaladoddi2019combined, lu2022rayleigh}.
In 
figure~\ref{fig:codevalidation},
we  compare our numerical results with prior experimental and two-dimensional numerical data~\citep{beckermann1989effect}. The demonstrated agreement between the three datasets validates our numerical implementation.

Moreover, the numerical data in figure \ref{fig:codevalidation}b further validate our derived melting curves and the transition criterion. This indicates that our results are applicable not only to the organic PCM n-octadecane, as discussed in the main article, but also to metallic materials (${\Pr}=0.0208$ herein corresponding to Gallium \citep{beckermann1989effect}).

\section{A Posterior Justification of Neglecting Advection Term} \label{Appendix Small Pelect number}
\begin{figure}
  \centerline{\includegraphics{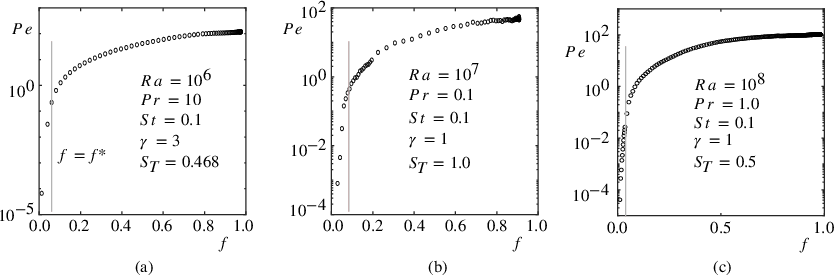}}
  \caption{ {Evolution of P\'eclet number in three configurations: (a) 
${\Ra}=1.0 \times 10^6$, ${\Pr}=10.0$, ${\St}=0.1$, $\gamma=3$, and $\ST=0.468$;  (b) ${\Ra}=1.0\times 10^7$, ${\Pr}=0.1$, ${\St}=0.1$, $\gamma=1$, and $\ST=1 $; (c)    ${\Ra}=1.0\times 10^8$, ${\Pr}=1.0$, ${\St}=0.1$, $\gamma=1$, and $\ST=0.5$. The vertical lines represent the critical liquid fraction $\fc$.
These configurations match  those in figure \ref{fig:validation}g--i. }}
\label{fig:peclet number}
\end{figure}

{We neglect the advection terms in \eqref{perturbation equations} to simply the stability analysis, assuming a low P\'eclect number ($\Pe = U_0\tilde w_b/\kappa$) in the conduction-dominated regime. This assumption is validated by examining the evolution of $\Pe$ in melting processes of three configurations, as shown in figure \ref{fig:peclet number}.
Within the conduction-dominated regime ($\f<\fc$), 
$\Pe$ remains low, reaching critical values of 0.22, 0.44, and 0.027 when $\f =\fc$. This directly justifies our low-$\Pe$ assumption. Furthermore, 
the predictive capacity (see figure \ref{fig:validation}g--i) of the theoretical criterion \eqref{transition criterion} derived from the simplified perturbation equations indirectly corroborates the neglect of advection terms.}

\begin{figure}
  \centerline{\includegraphics{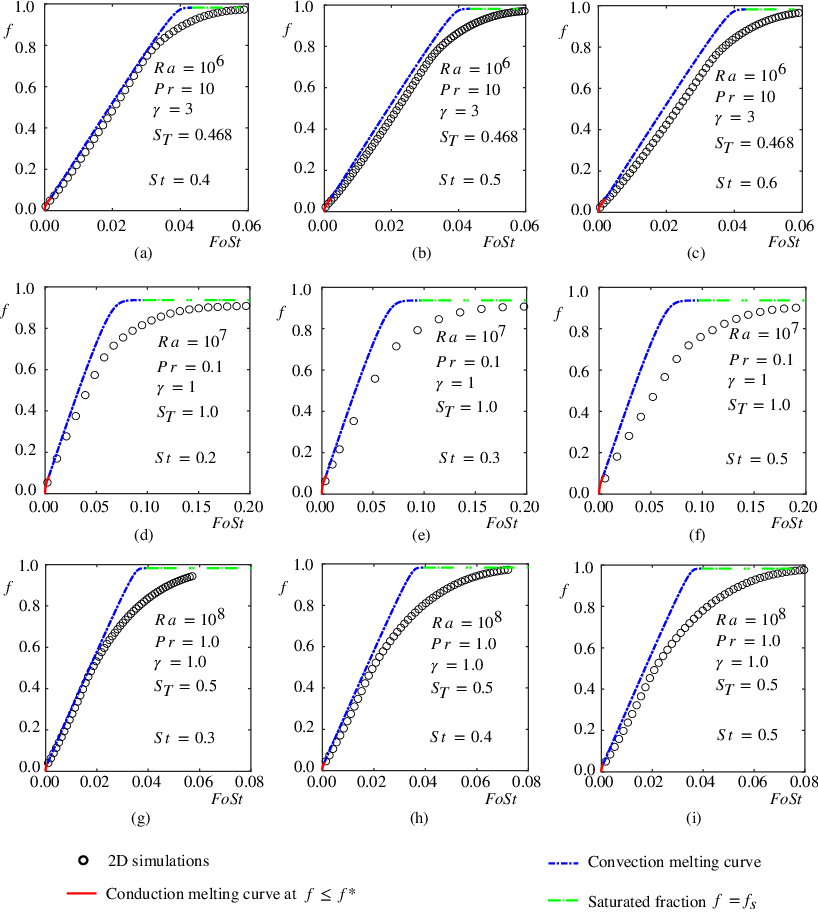}}
  \caption{Effects of the Stefan number on the validity of the model, illustrated for different configurations: : (a--c) ${\Ra}=10^6$, ${\Pr}=10$, $\gamma =3$, and $\ST=0.468$; (d--f)  ${\Ra}=10^7$, ${\Pr}=0.1$, $\gamma =1$, and $\ST=1$; (g--i) ${\Ra}=10^8$, ${\Pr}=1.0$, $\gamma =1$, and $\ST=0.5$. }
\label{fig:Ste effects}
\end{figure}

{
\section{Effect of Stefan Number on the Accuracy of the Theoretical Model} \label{Appendix Stefan's effect}
As discussed in the main text, our theory assumes $\St \ll 1$, 
so its accuracy depends on the value of $\St$. 
To investigate this dependence, we conduct a series of simulations, as depicted in Figure \ref{fig:Ste effects}.
As expected, the discrepancy between theoretical and numerical predictions increases with growing $\St$ for different parametric combinations.
Notably,   even though the maximum relative error reaches $20\%$, an overall acceptable agreement is seen in figure \ref{fig:Ste effects} a, d and g, where $\St = 0.2 \sim 0.4$. Thus, we may conclude that the model is valid for $\St \lessapprox \lp 0.2\sim0.4 \rp$. Indeed,
most $\St$ values in previous studies   \citep{beckermann1989effect,ho2013experimental,arasu2012numerical,kean2019numerical,regin2009analysis,khan2019thermodynamic} fall within this range, suggesting the relevance of our theoretical model.
}

\section{Melting of Nanoparticle-Enhanced PCM in A Basal Heating Cavity} \label{NEPCM melting}
Here, we apply the theoretical framework to a basal heating configuration, which differs from the lateral heating configuration in two ways. First, the isothermal heat source now serves as the cavity's bottom wall, instead of the vertical wall. Second, the storage materials are now nanoparticle-enhanced PCM (NEPCM), rather than pure PCM. 
The analytical solution derived from the theoretical framework successfully predicts the performance of different NEPCMs, despite featuring slightly different material properties. This demonstrates that the framework can accurately evaluate the effects of material properties. With this verified solution, we further provide a new explanation for how nanoparticles affect the melting rate.

\subsection{Analytical Solution}
\begin{figure}
  \centerline{\includegraphics{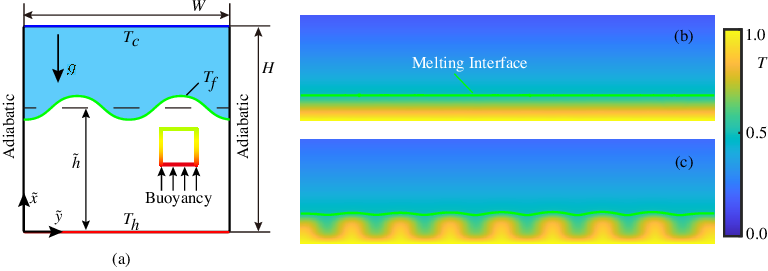}}
  \caption{(a) A rectangular cavity filled with subcooled PCM, with the temperature maintained at $T_h$ (hot) and $T_c$ (cold) at the bottom and top boundaries, respectively. (b) and (c) Typical temperature fields before and after the onset of convection, simulated at $\Ra=10^6$, $\Pr=10$, $\St=0.1$, $\ST=1.0$, and $\gamma=2.0$. Only the bottom halves of the fields are presented here.}
\label{Basal model}
\end{figure}

The basal heating configuration, shown in figure \ref{Basal model}a, can be obtained by simply rotating the lateral heating configuration by 90 degrees in the counter-clockwise direction. This rotation does not alter the energy equation or the dimensionless numbers, as repeated below for completeness,

\begin{equation}
\frac{\partial{\tilde T}}{\partial  \tilde t}+\boldsymbol{\tilde u} \cdot \grad    \tilde{T} =\kappa {\nabla^2}\tilde{T} -\frac{\mathcal{L}}{C_p}\frac{\partial {\f_{loc}}}{\partial \tilde t},
\label{energy equation for the whole cavity Appendix}
\end{equation}
\begin{equation}
{\Ra}=\frac{g\alpha \Delta T_l H^3}{\nu\kappa},~ {\Pr}=\frac{\nu}{\kappa},~\St=\frac{ C_p \Delta T_l}{\mathcal{L}}, 
~\ST=\frac{\Delta T_s}{\Delta T_l},~\gamma=\frac{W}{H}.
\label{control parameters Appendix}
\end{equation}

However, the rotation changes the thermal boundary conditions, thus leading to a different  melting behavior. Consequently, the Nusselt number $\Nu$, 
Fourier time ${\Fo}$, liquid fraction $\f$, and effective Rayleigh number $\Ra_e$ are redefined as 
\begin{subequations}\label{response parameters in basal}
\begin{align}
{\Nu}&=-\frac{\tilde h}{\Delta T_l}\langle \frac{\partial \tilde T}{\partial \tilde x} \rangle_{\tilde x=0},\label{eq:Nu}
\\
{\Fo}&=\frac{\kappa \tilde t}{H^2},\\
\f&=\frac{\tilde h}{H},\\
\Ra_e&=\frac{g\alpha \Delta T_l \tilde h^3}{\nu\kappa}=\Ra \f^3\label{eq:Rae},
\end{align}
\end{subequations}
where $\tilde h$  is the average height of the liquid layer,  and $\langle~ \rangle_{\tilde x=0}$ denotes spatial averaging along the hot wall at $\tilde x = 0$ (figure \ref{Basal model}a).

The transition criterion to differentiate  melting regimes in this configuration can be obtained through linear stability analysis. The base flow is
$\tilde{\boldsymbol{u}}_b = \mathbf{0}$, 
since the  buoyancy  is now horizontally homogeneous as indicated by $\partial \tilde T /\partial \tilde y =0$ (figure \ref{Basal model}a). 
Consequently, the instability phenomenon is very similar to that of the canonical Rayleigh-B\'ernard convection (figure \ref{Basal model}b and c), and has been extensively investigated \citep{davis1984pattern,kim2008onset, vasil2011dynamic,esfahani2018basal,madruga2018dynamic,favier2019rayleigh,purseed2020bistability}. 

For the present configuration, the transition criterion is described by a critical effective Rayleigh number $\Rac$. It ranges from 1708 to 1493~\citep{davis1984pattern}, depending on the relative heights of the liquid and solid layers when the melting process is saturated. The highest and lowest $\Rac$ correspond to similar liquid fractions, as indicated by $\lp 1708/1493\rp^{1/3}= 1.05$, following ~\eqref{eq:Rae}. 
Thus, the critical liquid fraction $\fc$ in subsequent theoretical modeling can be approximated by the median value, analogous to the lateral heating configuration
 \begin{equation}
\fc=\left(\frac{1600.5}{{\Ra}}\right)^{1/3}.
\label{transition criterion in basal}
\end{equation}

With this transition criterion $\fc$, we proceed to develop the mathematical model for the melting curve, following the same framework as in the lateral heating configuration. Neglecting the first term of ~\eqref{energy equation for the whole cavity Appendix}
 and applying Gauss's theorem to the remaining terms, we obtain
\begin{equation}
\frac{\mathcal{L}}{C_p}\frac{\partial \f}{\partial \tilde t} H =
\kappa \left[\langle\frac{\partial \tilde T}{\partial \tilde x}\rangle_{\tilde x=H} - \langle\frac{\partial \tilde T}{\partial \tilde x}\rangle_{\tilde x=0} \right].
\label{fl equation in basal}
\end{equation}
For the same reasons as in the lateral heating configuration, the temperature gradient at the cold wall can be considered steady, as
\begin{equation}
\langle\frac{\partial \tilde T}{\partial \tilde x}\rangle_{\tilde x=H}=-\frac{ \Delta T_s}{H-\tilde {h}},
\label{T gradient at cold wall in basal}
\end{equation}
where $H-\tilde h$ quantifies the average height of the solid phase (see  figure \ref{Basal model}a).

At the hot wall, the temperature gradient $\langle \partial \tilde T/ \partial \tilde x \rangle_{\tilde x=0}$  depends on the conduction and convection states of the liquid phase. When conduction dominates, this gradient becomes $-\langle \partial \tilde T/\partial \tilde x\rangle_{\tilde x=0}=\Delta T_l/\tilde h$, which  combining ~\eqref{fl equation in basal} and \eqref{T gradient at cold wall in basal}  yields the conduction melting curve
\begin{equation}
 \frac{\f(1+\ST)(\f-2\ST+\f\ST)-2\ST\ln[1-\f(1+\ST)]}{2(1+\ST)^3} 
= \Fo\St.
\label{conduction fl in basal}
\end{equation}
This curve exactly matches that in the lateral heating configurations, since the conduction heat transfer  solely depends on molecular motion, which is irrelevant to the orientation of the heat source.

As convection dominates, the temperature gradient can be modelled as $-\langle \partial \tilde T/\partial \tilde x \rangle_{\tilde x=0}={\Nu} \Delta T_l/ \tilde h$ following ~\eqref{eq:Nu}, which rearranges ~\eqref{fl equation in basal} to
\begin{equation}
\frac{\partial \f}{\partial \tilde t}  = \St \frac{\kappa}{H^2}
\left[  \frac{\Nu}{\f} -\frac{\ST}{1-\f} \right].
\label{fl equation in basal for temporary}
\end{equation}
The Nusselt number in the present configuration is $\Nu \approx 0.1 \Ra_e^{1/3} = 0.1 \Ra^{1/3} \f$~\citep{favier2019rayleigh}, a typical result predicted by the Grossmann-Lohse theory \citep{ahlers2009heat}. Substituting this correlation into ~\eqref{fl equation in basal for temporary},  we obtain the convection melting curve:
\begin{equation}
\frac{\f \mathcal{F} -\ST \ln(\f\mathcal{F}+\ST-\mathcal{F})}{\mathcal{F}^2} = \Fo \St +C,
\label{convection fl in basal}
\end{equation}
where $\mathcal{F} =0.1\Ra^{1/3}$ and $C$ is a constant to be determined. This curve differs from that in the lateral heating configuration, because the convection heat transfer depends on fluid flow, which in turn is influenced by the orientation of the heat source.

Because both  conduction and convection melting curves concatenate at $\left( {\Fo}^{\ast} \St,\fc \right)$,
subtracting ~\eqref{conduction fl in basal} from ~\eqref{convection fl in basal} yields the constant\begin{equation}
C= \frac{\fc \mathcal{F} -\ST \ln(\fc\mathcal{F}+\ST-\mathcal{F})}{\mathcal{F}^2}
-\frac{\fc(1+\ST)(\fc-2\ST+\fc\ST)-2\ST\ln[1-\fc(1+\ST)]}{2(1+\ST)^3}.
\label{C in basal}
\end{equation}
By  substituting ~\eqref{C in basal} into ~\eqref{convection fl in basal},  the domain of the logarithmic function necessitates 
\begin{equation}
\f < \f_s=1-\frac{\ST}{\mathcal F}.
\label{definition domain in lateral}
\end{equation}
This indicates that  melting will cease at a saturated liquid fraction $\f_s$, leaving  a portion of solid PCM  unmelted.

For now, we have derived the analytical solutions for the basal heating configuration, using the theoretical framework outlined in the main article for the lateral heating configuration. These solutions are independent of the cavity's aspect ratio $\gamma$, whereas those for the lateral configuration depend on $\gamma$. In the following, we will verify the new solutions by the results of \citet{kean2019numerical}.

\subsection{Verification of the Analytical Solutions}

The NEPCM used in \citet{kean2019numerical} consists of paraffin wax and various concentrations of ${\rm Al}_2{\rm O}_3$ nanoparticles, which can address the poor thermal conductivity of traditional PCMs \citep{kibria2015review,levin2013numerical,jebasingh2020comprehensive,yang2020thermophysical,tariq2020nanoparticles}. 
The thermophysical properties of paraffin wax, ${\rm Al}_2{\rm O}_3$ nanoparticles, and NEPCM are summarized in Table~\ref{tab:properties}. Here, an experimentally fitted model is utilized to calculate the NEPCM's  density $\rho$, specific heat $C_p$, latent heat $\mathcal{L}$, dynamic viscosity $\mu$, and thermal conductivity $K$~\citep{chow1996thermal,vajjha2010numerical}, as

\begin{table}
  \begin{center}
\def~{\hphantom{0}}
  \begin{tabular}{lcccc}
      Property& Paraffin wax&   ${\rm Al}_2{\rm O}_3$ & Wax + $2\%$ ${\rm Al}_2{\rm O}_3$&Wax + $5\%$ ${\rm Al}_2{\rm O}_3$\\[3pt]
       Density ($\rm Kg/m^3$)& $\frac{750}{0.001(\tilde T-319.15)+1}$& 3600& 802.36&888.00\\
       Specific heat ($\rm J/ (KgK)$)& 2890& 765& 2699.31&2459.26\\
       Latent heat ($\rm J/Kg$)& 173400& --& 157839.94&138251.52\\
       Thermal conductivity ($\rm W/mK$)& 0.12& 36& 0.16&0.17\\
       Dynamic viscosity ($\rm Kg/ms$)& $0.001\exp(1790/\tilde T-4.25)$& --& 0.0044&0.0066\\
 Fusion temperature ($\rm K$)& 321& --& --&--\\
 Thermal expansion ($\rm 1/K$)& $10^{-3}$& --& --&--\\
  \end{tabular}
  \caption{Properties of paraffin wax, ${\rm Al}_2{\rm O}_3$ nanoparticles,  and two different NEPCMs  \citep{arasu2012numerical,kean2019numerical}.} 
  \label{tab:properties}
  \end{center}
\end{table}

\begin{subequations}\label{NEPCM properties}
\begin{align}
\rho & = \phiv \rho_{np} +(1-\phiv)\rho_{pw}, \\
C_p & =\frac{\phiv(\rho c_p)_{np} + (1-\phiv)(\rho C_p)_{pw}}{\rho}, \\
\mathcal{L} & =\frac{(1-\phiv)(\rho \mathcal L)_{pw}}{\rho}, \\
\mu & =0.983\exp(12.959\phiv)\mu_{pw}, \\
K & =\frac{K_{np}+2K_{pw}-2(K_{pw}-K_{np})\phiv}{K_{np}+2K_{pw}+(K_{pw}-K_{np})\phiv}K_{pw} + 5\times 10^4 \beta_k \zeta \phiv (\rho C_p)_{pw} \sqrt{\frac{1.381\times 10^{-23} \tilde T}{\rho_{np} d_{np}}} m, \label{eq:K}
\end{align}    
\end{subequations}
where $\phiv$ represents the volume fraction of nanoparticles; here, the subscripts `${\it pw}$' and `${\it np}$' denote paraffin wax and ${\rm Al}_2{\rm O}_3$ nanoparticles, respectively. 
In ~\eqref{eq:K}, $d_{np}$ denotes the nanoparticle diameter, and $\zeta$ is a correction factor  equal to  the local liquid fraction, and $\beta_k$ and $m$ are expressed by
\begin{align}
\beta_k & =8.4407(100\phiv)^{-1.07304}, \nonumber \\
m & =(2.8217\times 10^{-2}\phiv + 3.917\times10^{-3})\frac{T}{273} -(3.0669\times10^{-2}\phiv + 3.91123\times10^{-3}).
\label{subparameters in NEPCM properties}
\end{align}

In \citet{kean2019numerical}, the cavity dimensions are $W=H=0.025~{\rm m}$, and $d_{np}=59~{\rm nm}$.  The temperatures at hot and cold walls are $T_h=330~{\rm K}$ and $T_c=300~{\rm K}$, respectively. The properties of the NEPCM depend on the temperature field $\tilde T$, which can be obtained by simulations but poses a challenge in model calculations. Consequently, we choose the reference temperature $T_0=(T_h+T_f)/2$ as the characteristic value of $\tilde T$. At this characteristic temperature, the properties of the NEPCM can be calculated using  \eqref{NEPCM properties} and \eqref{subparameters in NEPCM properties}, as listed in  Table~\ref{tab:properties}. 
Notably, other characteristic temperatures, such as $T_h$ or $T_f$, will not significantly affect the results.

\begin{figure}
  \centerline{\includegraphics{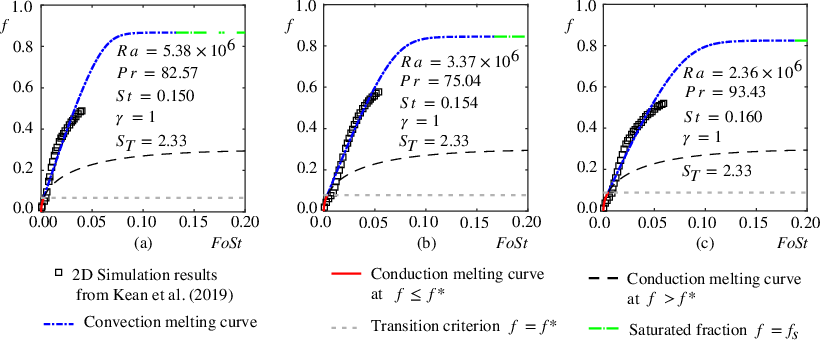}}
  \caption{Comparison between our theoretical  prediction and  numerical results (squares) of 
  \citet{kean2019numerical}  for a cavity filled with: (a) Pure paraffin wax, (b) NEPCM with $2\%$ ${\rm Al}_2{\rm O}_3$, and (c) NEPCM with $5\%$ ${\rm Al}_2{\rm O}_3$.}
\label{fig:validation_basal}
\end{figure}

With the properties listed in  Table \ref{tab:properties},  the dimensionless numbers in   \eqref{control parameters Appendix} can be calculated for pure paraffin wax, NEPCM with $2\%$ ${\rm Al}_2{\rm O}_3$, and NEPCM with $5\%$ ${\rm Al}_2{\rm O}_3$.  The resulting three sets of numbers can then be substituted into  the analytical solutions, \eqref {transition criterion in basal}, \eqref{conduction fl in basal},  \eqref{convection fl in basal}, and \eqref{definition domain in lateral},  to generate three melting curves. As shown in figure \ref{fig:validation_basal}, these curves agree with the numerical data of \citet{kean2019numerical},  thereby verifying our theoretical solutions.
With these verified solutions, we will further discuss the effects of nanoparticles on the melting rate.

\subsection{Effects of Nanoparticles on Melting Rate}

According to  \eqref{convection fl in basal},   the dimensionless melting times   $\Fo \St$  to achieve  $\f=0.5$ are 0.0332, 0.0399, and 0.0462 for pure paraffin wax, NEPCM with $2\%$ ${\rm Al}_2{\rm O}_3$, and NEPCM with $5\%$ ${\rm Al}_2{\rm O}_3$, respectively.  Correspondingly, the dimensional times  $\tilde t = \Fo H^2/\kappa $ are 2480.3 s, 2197.9 s, and 2285.7 s, confirming the numerically revealed guideline that  NEPCM with  $2\%$ ${\rm Al}_2{\rm O}_3$  exhibits the optimal melting rate ~\citep{kean2019numerical}.  This demonstrates that although  pure paraffin wax, NEPCM with $2\%$ ${\rm Al}_2{\rm O}_3$, and NEPCM with $5\%$ ${\rm Al}_2{\rm O}_3$  have slightly different properties (Table \ref{tab:properties}), their performance can be accurately differentiated by the present model.

The nonmonotonic dependence of the melting rate on $\phiv$ is widely attributed to the competition between $\mu$ and $K$ \citep{ho2008numerical,ho2010natural,ho2013experimental,arasu2012numerical,kibria2015review}. 
When $\phiv$ increases, both $K$ and $\mu$ of NEPCM are enlarged. 
These studies argued that a larger $K$ enhances heat transfer thus benefits melting, while an augmented $\mu$ hinders heat convection and hence degenerates melting. 
These opposing effects collectively lead to a nonmonotonic variation in melting rate  with $\phiv$. Despite its general acceptance, this explanation might lack rigor. According to ${\Ra}={g\alpha \Delta T_l H^3}/{(\nu\kappa)}={g\alpha \Delta T_l H^3 \rho^2 C_p}/{(\mu K)}$,  increasing $K$ and $\mu$ both decrease $\Ra$, thereby diminishing convection and negatively impacting melting. 
This indicates that the effects of $K$ and $\mu$ are not always antagonistic, contrary to the reported explanation. Additionally, nanoparticles not only affect  $K$ and $\mu$  but also change $\rho$, $C_p$, and $\mathcal L$, as illustrated in  Table \ref{tab:properties}. Thus, the reported explanation, which only considers $K$ and $\mu$ in competition,  might not fully elucidate the impacts of nanoparticles on the melting rate.

On the other hand, our model offers a more rigorous explanation for the nonmonotonic relation. Our analytical solution, \eqref{conduction fl in basal} and \eqref {convection fl in basal}, shows that the melting rate depends on three dimensionless parameters, $\Ra$, $\St$, and $\ST$. Because the fusion temperature $T_f$ barely changes with $\phiv$~\citep{yang2020thermophysical}, the impact of $\ST$  can be safely neglected, allowing us to focus on $\Ra$ and $\St$ only. As depicted in figure~\ref{fig:validation_basal}, increasing $\phiv$ decreases $\Ra$ and increases $\St$ simultaneously. A lower $\Ra$ weakens convection, thereby impeding melting. Conversely, a higher  $\St=C_p\Delta T_l/\mathcal L$ signifies a lower latent heat, hence accelerating melting. These two contrasting trends thus result in the nonmonotonic dependence of the melting rate on $\phiv$. The expressions for $\Ra$ and $\St$ include all the properties affected by nanoparticles, \ie $K$, $\mu$, $\rho$, $C_p$, and $\mathcal L$. Hence, compared to the reported explanation, our explanation based on $\Ra$ and $\St$ encodes a more comprehensive account of how nanoparticles influence the melting rate.

\section{Melting of PCM in An Inclined Cavity} \label{inclined cavity}

This section illustrates the application of the theoretical framework in the configuration of an inclined cavity. Compared to the lateral heating configuration, this configuration has a distinct geometric feature characterized by the inclined angle. Moreover, the heat source in this configuration operates at a constant power, rather than maintaining an isothermal temperature.
Despite these differences, the melting curve for the current configuration can be derived using the same theoretical framework.
Thus, it is reasonable to claim that the theoretical framework is capable of considering the effects of various geometric features and operational conditions. Using the resulting curve, we further calculate
the storage energy, which is consistent with the experimental data  \citep{korti2020experimental}, confirming the effectiveness of the framework in predicting the overall system performance.

\begin{figure}
  \centerline{\includegraphics{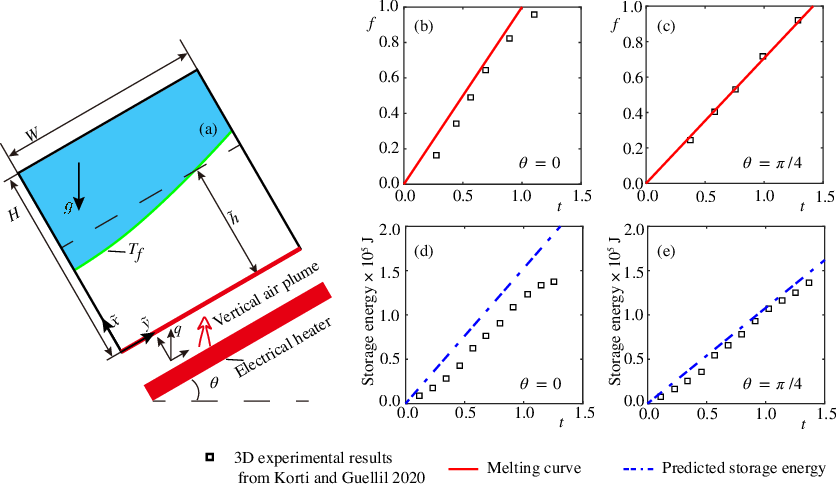}}
  \caption{(a) An inclined cavity filled with subcooled PCM, with the bottom wall heated by a parallel electrical heater. 
 (b) and (c) Comparison between  the melting curve and the experimental results~\citep{korti2020experimental}, when the inclination angle $\theta=0$ and $\pi/4$, respectively. (d) and (e), similar to (b) and (c), but for the storage energy.}
\label{Inclined model}
\end{figure}

The inclined cavity investigated here  matches that in \citet{korti2020experimental}, as shown in figure \ref{Inclined model}a. Such an inclined setting is typically encountered when PCM containers are deployed on the non-flat surface of a working device \citep{khan2024review}.  
The dimensions of this cavity are $L=0.045 ~{\rm m}$ in depth, $W=0.12 ~{\rm m}$ in width, and $H=0.12~{\rm m}$ in height. All its walls are insulated except for the bottom one, which receives heat from a parallel electrical heater operating at a constant power of $q=2700~{\rm W/m^2}$.  These thermal boundary conditions are frequently encountered in applications such as electronic cooling and solar energy collection. The entire system  can rotate by an angle $\theta$  (figure \ref{Inclined model}a), and maintains this orientation. The PCM (paraffin wax, CAS 8002-74-2) is initially in the solid phase at a temperature of $T_c=296.65 ~{\rm K}$. The thermophysical properties of the PCM are summarized in Table \ref{tab:Inclied PCM properties}.  

\begin{table}
  \begin{center}
\def~{\hphantom{0}}
  \begin{tabular}{lc}
      Property& Value\\[3pt]
       Density ($\rm Kg/m^3$)& 916\\
       Specific heat ($\rm J/ KgK$)& 2900\\
       Specific latent heat ($\rm J/Kg$)& 176000\\
       Thermal conductivity ($\rm W/mK$)& 0.12\\
       Dynamic viscosity ($\rm Kg/ms$)& 0.0036\\
 Fusion temperature ($\rm K$)& 324.65\\
 Thermal expansion ($\rm 1/K$)& $9.1\times 10^{-4}$\\
  \end{tabular}
  \caption{Properties of the paraffin wax, CAS 8002-74-2~\citep{korti2020experimental}.} 
  \label{tab:Inclied PCM properties}
  \end{center}
\end{table}


Using the same framework as applied to the lateral heating cavity, the melting curve for the inclined cavity can be derived as follows. By neglecting the sensible heat term in  \eqref{energy equation for the whole cavity} and applying Gauss's theorem to the remaining terms, we obtain:

\begin{equation}
\frac{\mathcal{L}}{C_p}\frac{\partial \f}{\partial \tilde t}W H =
 \int (\kappa \grad \tilde T)_{\tilde x=0} \cdot \boldsymbol{n} {\rm d}A.
\label{fl equation in inclined}
\end{equation}
The area element at the bottom wall is ${\rm d}A={\rm d}y\times 1$, with  $\boldsymbol{n}=(-1,~0)$ representing the outward unit normal vector (figure \ref{Inclined model}a). The temperature gradient at this wall $(\grad \tilde T)_{\tilde x=0} $ can be related to the heater's power $q$, as explained below.

The PCM container is separated from the electrical heater by a small gap, within which vertical air plumes arise due to buoyancy  (figure \ref{Inclined model}a).  
Receiving it from the heater, the plumes transfer the heat to the cavity when they impinge on the bottom wall. 
Thus, during the impingement process, we obtain:
\begin{equation}
-\rho C_p \kappa (\grad \tilde T)_{\tilde x=0}=q(\cos\theta,~\sin \theta),
\label{Bottom heat flux}
\end{equation}
where the right-hand side represents the heat flux carried by the plumes, as shown in  figure \ref{Inclined model}a. This equation indicates that the wall temperature gradient does not depend on the conduction and convection melting regimes. 
Therefore, linear stability analysis, which is required in the lateral heating configuration to distinguish different regimes, is not necessary in this configuration.

Based on  \eqref{Bottom heat flux}, the integration of  wall temperature gradient can be calculated as
\begin{equation}
 \int (\kappa \grad \tilde T)_{\tilde x=0} \cdot \boldsymbol{n} {\rm d}A
 =  \int_0^W \frac{q}{\rho C_p} \cos\theta {\rm d}y
 =\frac{q\cos\theta}{\rho C_p}W.
\label{wall T gradient in inclined cavity}
\end{equation}

Subsequently, the melting curve can be obtained from  \eqref{fl equation in inclined}, as 
\begin{equation}
\f=t\cos \theta,
\label{Inclined melting curve}
\end{equation}
where the dimensionless time is $t=q \tilde t/(\rho H \mathcal L)$. This curve does not impose any constraint on the liquid fraction $\f$. Thus, its saturated value is $\f_s =1$, indicating that the solid PCM can be completely melted. 
Notably,  at a large inclination, e.g., $\theta\geq \pi/2$,  the vertical air plumes do not impinge on the bottom wall. Accordingly,  \eqref{Bottom heat flux} no longer holds, 
and determining $ (\grad \tilde T)_{\tilde x=0}$ would require analysis of the flow field within the gap. This is beyond the scope of this study, which is hence not pursued here.

According to   \eqref{Inclined melting curve}, the derived melting curve  accounts for the effects of material properties ($\rho$ and $\mathcal L$), geometric features ($H$ and $\theta$), and operational condition ($q$). This curve is well-verified against the experimental results~\citep {korti2020experimental}  at $\theta =0$ and $\theta =\pi/4$, as shown in in figures \ref{Inclined model}b and c. 
This verified curve can then be used to calculate the energy density \eqref{PCMs indicators}.~  
The energy density $\ED$, when multiplied by the container volume  ($W\times H \times L$), yields the storage energy, which is consistent with the experimental result (figures \ref{Inclined model}d and e), thereby confirming the model's effectiveness.
The power density is the time derivative of $\ED$, which is hence not verified here separately.

\section{Melting of PCM in An Annular Tube} \label{Annular Tube}

Here, we apply the theoretical framework to the configuration of an annular tube. This configuration differs significantly from the lateral heating configuration in terms of the container's geometry. Moreover, the heat source in this configuration operates at constant power, rather than maintaining an isothermal temperature.
Despite these differences, the melting curve for the current configuration can be derived using the same theoretical framework. This curve is well validated by the experimental results \citep{dhaidan2013experimental}, confirming the capability of the theoretical framework to account for the effects of various geometric features and operational conditions.

\begin{figure}
  \centerline{\includegraphics{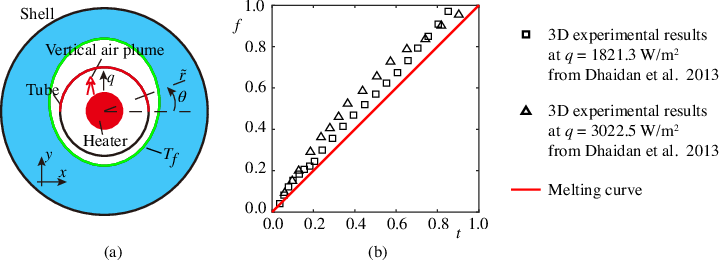}}
  \caption{(a) An annular tube filled with subcooled PCM. 
 (b)  Comparison between our predicted melting curve and the experimental results  \citep{dhaidan2013experimental}. }
\label{annular model}
\end{figure}

The configuration of the annular tube is the same as that in \citet{dhaidan2013experimental}, as shown in figure \ref{annular model}a.
In this setup, the outer shell and inner tube are coaxially arranged with radii of of $R_o=22.25 ~{\rm mm}$ and $R_i = 9.5 ~{\rm mm }$, respectively.  The outer shell is insulated, while the inner tube is heated by an electrical heater operating at a constant power $q$. The space between the shell and tube contains n-octadecane paraffin. Initially, this PCM is in the solid phase at a temperature of  $T_c=295.85 ~{\rm K}$, and its thermophysical properties are summarized in Table \ref{tab:annular PCM properties}. 

\begin{table}
  \begin{center}
\def~{\hphantom{0}}
  \begin{tabular}{lc}
      Property& Value\\[3pt]
       Density ($\rm Kg/m^3$)& 770\\
       Specific heat ($\rm J/ KgK$)& 2196\\
       Specific latent heat ($\rm J/Kg$)& 243500\\
       Thermal conductivity ($\rm W/mK$)& 0.148\\
       Kinematic viscosity ($\rm m^2/s$)& $5\times 10^{-6}$\\
 Fusion temperature ($\rm K$)& 301.15\\
 Thermal expansion ($\rm 1/K$)& $9.1\times 10^{-4}$\\
  \end{tabular}
  \caption{Properties of n-octadecane paraffin \citep{dhaidan2013experimental}.} 
  \label{tab:annular PCM properties}
  \end{center}
\end{table}

Following the same framework outlined in the main article, the melting curve for this annular tube is derived as follows.
By neglecting the sensible heat term in   \eqref{energy equation for the whole cavity}  and applying Gauss's theorem to the remaining terms, we obtain:

\begin{equation}
\frac{\mathcal{L}}{C_p}\frac{\partial \f}{\partial \tilde t}(\pi R_o^2-R_i^2) =
 \int (\kappa \grad \tilde T)_{\tilde r=R_i} \cdot \boldsymbol{n} {\rm d}A.
\label{fl equation in annular}
\end{equation}
The area element of the tube wall  is ${\rm d}A={\rm d}\theta  R_i \times 1$, with  $\boldsymbol{n}=(-\cos\theta,~-\sin\theta)$ representing the outward unit normal vector (figure \ref{annular model}a). The temperature gradient at this wall   $( \grad \tilde T)_{\tilde r=R_i} $ can be related to the heater's power $q$, as shown below.

The inner tube is separated from the electrical heater by a small gap, within which vertical air plumes arise due to  buoyancy (figure \ref{annular model}a). These plumes receive heat from the heater, and then transfer it to the upper half of the inner tube when they impinge on the tube wall. Thus, during the impingement process, we obtain:
\begin{equation}
-\rho C_p \kappa (\grad \tilde T)_{\tilde r=R_i}=q(0,~1),
\label{Bottom heat flux in annular}
\end{equation}
where the right-hand side represents the heat flux carried by the plumes, as shown in  figure \ref{annular model}a. This equation indicates that the wall temperature gradient does not depend on  the conduction and convection melting regimes. 
Therefore, linear stability analysis is not necessary here.\\

Using  \eqref{Bottom heat flux in annular}, the integration of  wall temperature gradient can be calculated by
\begin{equation}
 \int (\kappa \grad \tilde T)_{\tilde r=R_i} \cdot \boldsymbol{n} {\rm d}A
 =  \int_0^\pi \frac{q}{\rho C_p} \sin\theta~R_i {\rm d}\theta
 =\frac{2 R_i q}{\rho C_p}.
\label{wall T gradient in annular cavity}
\end{equation}
Subsequently, the melting curve can be obtained from  \eqref{fl equation in annular} as 
\begin{equation}
\f=t,
\label{annular melting curve}
\end{equation}
where the dimensionless time is $t=2R_i q \tilde t/[\pi \rho \mathcal L (R_o^2-R_i^2)]$.  This curve does not impose any constraint on $\f$. Thus, the saturated value should be $\f_s =1$, indicating that the solid PCM can be completely melted.

According to  \eqref{annular melting curve}, the derived melting curve accounts for the effects of material properties ($\rho$ and $\mathcal L$), geometric features ($R_i$ and $R_o$), and operational condition ($q$). 
Figure~\ref{annular model}b suggests that
this curve agrees reasonably well with the experimental data~\citep {dhaidan2013experimental}  when $q =1821.3 ~{\rm W/m^2}$ and $q =3022.5 ~{\rm W/m^2}$, verifying our theoretical solutions.


\begin{thebibliography}{76}
\expandafter\ifx\csname natexlab\endcsname\relax\def\natexlab#1{#1}\fi
\def\au#1{#1} \def\ed#1{#1} \def\yr#1{#1}\def\at#1{#1}\def\jt#1{\textit{#1}}
  \def\bt#1{#1}\def\bvol#1{\textbf{#1}} \def\vol#1{#1} \def\pg#1{#1}
  \def\publ#1{#1}\def\arxiv#1{#1}\def\org#1{#1}\def\st#1{\textit{#1}}

\bibitem[Ahlers {\em et~al.\/}(2022)Ahlers, Bodenschatz, Hartmann, He, Lohse,
  Reiter, Stevens, Verzicco, Wedi, Weiss {\em et~al.\/}]{ahlers2022aspect}
{\sc \au{Ahlers, G.}, \au{Bodenschatz, E.}, \au{Hartmann, R.}, \au{He, X.},
  \au{Lohse, D.}, \au{Reiter, P.}, \au{Stevens, R. J. A.~M.}, \au{Verzicco,
  R.}, \au{Wedi, M.}, \au{Weiss, S.} \& \au{others}} \yr{2022}  \at{Aspect
  ratio dependence of heat transfer in a cylindrical {R}ayleigh-{B}{\'e}nard
  cell}.  \jt{Phys. Rev. Lett.}  \bvol{128}~(8),  \pg{084501}.

\bibitem[Ahlers {\em et~al.\/}(2009)Ahlers, Grossmann \& Lohse]{ahlers2009heat}
{\sc \au{Ahlers, G.}, \au{Grossmann, S.} \& \au{Lohse, D.}} \yr{2009}  \at{Heat
  transfer and large scale dynamics in turbulent {R}ayleigh-{B}{\'e}nard
  convection}.  \jt{Rev. Mod. Phys.}  \bvol{81}~(2),  \pg{503}.

\bibitem[Arasu \& Mujumdar(2012)]{arasu2012numerical}
{\sc \au{Arasu, A.~V.} \& \au{Mujumdar, A.~S.}} \yr{2012}  \at{Numerical study
  on melting of paraffin wax with {A}l$_2${O}$_3$ in a square enclosure}.
  \jt{Int. Commun. Heat Mass}  \bvol{39}~(1),  \pg{8--16}.

\bibitem[Azad {\em et~al.\/}(2021{\natexlab{{\em a\/}}})Azad, Groulx \&
  Donaldson]{azad2021naturalI}
{\sc \au{Azad, M.}, \au{Groulx, D.} \& \au{Donaldson, A.}}
  \yr{2021{\natexlab{{\em a\/}}}}  \at{Natural convection onset during melting
  of phase change materials: Part {I}--effects of the geometry, {S}tefan
  number, and degree of subcooling}.  \jt{Int. J. Therm. Sci.}  \bvol{170},
  \pg{107180}.

\bibitem[Azad {\em et~al.\/}(2021{\natexlab{{\em b\/}}})Azad, Groulx \&
  Donaldson]{azad2021natural}
{\sc \au{Azad, M.}, \au{Groulx, D.} \& \au{Donaldson, A.}}
  \yr{2021{\natexlab{{\em b\/}}}}  \at{Natural convection onset during melting
  of phase change materials: Part {II}--effects of {F}ourier, {G}rashof, and
  {R}ayleigh numbers}.  \jt{Int. J. Therm. Sci.}  \bvol{170},  \pg{107062}.

\bibitem[Azad {\em et~al.\/}(2022)Azad, Groulx \& Donaldson]{azad2022natural}
{\sc \au{Azad, M.}, \au{Groulx, D.} \& \au{Donaldson, A.}} \yr{2022}
  \at{Natural convection onset during melting of phase change materials: Part
  {III}--global correlations for onset conditions}.  \jt{Int. J. Therm. Sci.}
  \bvol{172},  \pg{107368}.

\bibitem[Beckermann \& Viskanta(1989)]{beckermann1989effect}
{\sc \au{Beckermann, C.} \& \au{Viskanta, R.}} \yr{1989}  \at{Effect of solid
  subcooling on natural convection melting of a pure metal}.  \jt{J. Heat
  Transfer}  \bvol{111},  \pg{416--424}.

\bibitem[Chen {\em et~al.\/}(2022)Chen, Liu, Gao \& Wang]{chen2022advanced}
{\sc \au{Chen, X.}, \au{Liu, P.}, \au{Gao, Y.} \& \au{Wang, G.}} \yr{2022}
  \at{Advanced pressure-upgraded dynamic phase change materials}.  \jt{Joule}
  \bvol{6}~(5),  \pg{953--955}.

\bibitem[Chow {\em et~al.\/}(1996)Chow, Zhong \& Beam]{chow1996thermal}
{\sc \au{Chow, L.~C.}, \au{Zhong, J.~K.} \& \au{Beam, J.~E.}} \yr{1996}
  \at{Thermal conductivity enhancement for phase change storage media}.
  \jt{Int. Commun. Heat Mass}  \bvol{23}~(1),  \pg{91--100}.

\bibitem[Davis {\em et~al.\/}(1984)Davis, M{\"u}ller \&
  Dietsche]{davis1984pattern}
{\sc \au{Davis, S.~H.}, \au{M{\"u}ller, U.} \& \au{Dietsche, C.}} \yr{1984}
  \at{Pattern selection in single-component systems coupling {B}{\'e}nard
  convection and solidification}.  \jt{J. Fluid Mech.}  \bvol{144},
  \pg{133--151}.

\bibitem[Dhaidan \& Khodadadi(2015)]{dhaidan2015melting}
{\sc \au{Dhaidan, N.~S.} \& \au{Khodadadi, J.~M.}} \yr{2015}  \at{Melting and
  convection of phase change materials in different shape containers: A
  review}.  \jt{Renew. Sust. Energ. Rev.}  \bvol{43},  \pg{449--477}.

\bibitem[Dhaidan {\em et~al.\/}(2013)Dhaidan, Khodadadi, Al-Hattab \&
  Al-Mashat]{dhaidan2013experimental}
{\sc \au{Dhaidan, N.~S.}, \au{Khodadadi, J.~M.}, \au{Al-Hattab, T.~A.} \&
  \au{Al-Mashat, S.~M.}} \yr{2013}  \at{Experimental and numerical
  investigation of melting of {N}e{PCM} inside an annular container under a
  constant heat flux including the effect of eccentricity}.  \jt{Int. J. Heat
  Mass Transf.}  \bvol{67},  \pg{455--468}.

\bibitem[Duan {\em et~al.\/}(2019)Duan, Xiong \& Yang]{duan2019melting}
{\sc \au{Duan, J.}, \au{Xiong, Y.} \& \au{Yang, D.}} \yr{2019}  \at{On the
  melting process of the phase change material in horizontal rectangular
  enclosures}.  \jt{Energies}  \bvol{12}~(16),  \pg{3100}.

\bibitem[Dutil {\em et~al.\/}(2011)Dutil, Rousse, Salah, Lassue \&
  Zalewski]{dutil2011review}
{\sc \au{Dutil, Y.}, \au{Rousse, D.~R.}, \au{Salah, N.~B.}, \au{Lassue, S.} \&
  \au{Zalewski, L.}} \yr{2011}  \at{A review on phase-change materials:
  Mathematical modeling and simulations}.  \jt{Renew. Sust. Energ. Rev.}
  \bvol{15}~(1),  \pg{112--130}.

\bibitem[Esfahani {\em et~al.\/}(2018)Esfahani, Hirata, Berti \&
  Calzavarini]{esfahani2018basal}
{\sc \au{Esfahani, B.~R.}, \au{Hirata, S.~C.}, \au{Berti, S.} \&
  \au{Calzavarini, E.}} \yr{2018}  \at{Basal melting driven by turbulent
  thermal convection}.  \jt{Phys. Rev. Fluids}  \bvol{3}~(5),  \pg{053501}.

\bibitem[Favier {\em et~al.\/}(2019)Favier, Purseed \&
  Duchemin]{favier2019rayleigh}
{\sc \au{Favier, B.}, \au{Purseed, J.} \& \au{Duchemin, L.}} \yr{2019}
  \at{{R}ayleigh-{B}{\'e}nard convection with a melting boundary}.  \jt{J.
  Fluid Mech.}  \bvol{858},  \pg{437--473}.

\bibitem[Feng {\em et~al.\/}(2007)Feng, Han \& Owen]{feng2007coupled}
{\sc \au{Feng, Y.~T.}, \au{Han, K.} \& \au{Owen, D. R.~J.}} \yr{2007}
  \at{Coupled lattice boltzmann method and discrete element modelling of
  particle transport in turbulent fluid flows: Computational issues}.  \jt{Int.
  J. Numer. Methods Eng.}  \bvol{72}~(9),  \pg{1111--1134}.

\bibitem[Fu {\em et~al.\/}(2022)Fu, Yan, Gurumukhi, Garimella, King \&
  Miljkovic]{fu2022high}
{\sc \au{Fu, W.}, \au{Yan, X.}, \au{Gurumukhi, Y.}, \au{Garimella, V.~S.},
  \au{King, W.~P.} \& \au{Miljkovic, N.}} \yr{2022}  \at{High power and energy
  density dynamic phase change materials using pressure-enhanced close contact
  melting}.  \jt{Nat. Energy}  \bvol{7}~(3),  \pg{270--280}.

\bibitem[Gao \& Lu(2021)]{gao2021machine}
{\sc \au{Gao, T.} \& \au{Lu, W.}} \yr{2021}  \at{Machine learning toward
  advanced energy storage devices and systems}.  \jt{i{S}cience}
  \bvol{24}~(1),  \pg{101936}.

\bibitem[Gau \& Viskanta(1986)]{gau1986melting}
{\sc \au{Gau, C.} \& \au{Viskanta, R.}} \yr{1986}  \at{Melting and
  solidification of a pure metal on a vertical wall}.  \jt{J. Heat Transfer}
  \bvol{108},  \pg{174--181}.

\bibitem[Gerkman \& Han(2020)]{gerkman2020toward}
{\sc \au{Gerkman, M.~A.} \& \au{Han, G. G.~D.}} \yr{2020}  \at{Toward
  controlled thermal energy storage and release in organic phase change
  materials}.  \jt{Joule}  \bvol{4}~(8),  \pg{1621--1625}.

\bibitem[Guo \& Shu(2013)]{guo2013lattice}
{\sc \au{Guo, Z.} \& \au{Shu, C.}} \yr{2013} {\em Lattice {B}oltzmann {M}ethod
  and Its Application in Engineering\/}, ,  \vol{vol.~3}.  \publ{World
  Scientific}.

\bibitem[Guo {\em et~al.\/}(2002)Guo, Zheng \& Shi]{Guo2002non}
{\sc \au{Guo, Z.-L.}, \au{Zheng, C.-G.} \& \au{Shi, B.-C.}} \yr{2002}
  \at{Non-equilibrium extrapolation method for velocity and pressure boundary
  conditions in the lattice {B}oltzmann method}.  \jt{Chinese Phys.}
  \bvol{11}~(4),  \pg{366}.

\bibitem[Gur {\em et~al.\/}(2012)Gur, Sawyer \& Prasher]{gur2012searching}
{\sc \au{Gur, I.}, \au{Sawyer, K.} \& \au{Prasher, R.}} \yr{2012}
  \at{Searching for a better thermal battery}.  \jt{Science}
  \bvol{335}~(6075),  \pg{1454--1455}.

\bibitem[He {\em et~al.\/}(2022)He, Guo \& Zhang]{he2022performance}
{\sc \au{He, Z.}, \au{Guo, W.} \& \au{Zhang, P.}} \yr{2022}  \at{Performance
  prediction, optimal design and operational control of thermal energy storage
  using artificial intelligence methods}.  \jt{Renew. Sust. Energ.}
  \bvol{156},  \pg{111977}.

\bibitem[Ho {\em et~al.\/}(2008)Ho, Chen \& Li]{ho2008numerical}
{\sc \au{Ho, C.-J.}, \au{Chen, M.~W.} \& \au{Li, Z.~W.}} \yr{2008}
  \at{Numerical simulation of natural convection of nanofluid in a square
  enclosure: effects due to uncertainties of viscosity and thermal
  conductivity}.  \jt{Int. J. Heat Mass Transf.}  \bvol{51}~(17-18),
  \pg{4506--4516}.

\bibitem[Ho \& Gao(2013)]{ho2013experimental}
{\sc \au{Ho, C.-J.} \& \au{Gao, J.~Y.}} \yr{2013}  \at{An experimental study on
  melting heat transfer of paraffin dispersed with {Al}$_2${O}$_3$
  nanoparticles in a vertical enclosure}.  \jt{Int. J. Heat Mass Transf.}
  \bvol{62},  \pg{2--8}.

\bibitem[Ho {\em et~al.\/}(2010)Ho, Liu, Chang \& Lin]{ho2010natural}
{\sc \au{Ho, C.~J.}, \au{Liu, W.~K.}, \au{Chang, Y.~S.} \& \au{Lin, C.~C.}}
  \yr{2010}  \at{Natural convection heat transfer of alumina-water nanofluid in
  vertical square enclosures: An experimental study}.  \jt{IInt. J. Therm.
  Sci.}  \bvol{49}~(8),  \pg{1345--1353}.

\bibitem[Ho \& Viskanta(1984)]{ho1984heat}
{\sc \au{Ho, C-J} \& \au{Viskanta, R}} \yr{1984}  \at{Heat transfer during
  melting from an isothermal vertical wall}.  \jt{J. Heat Transfer}
  \bvol{106},  \pg{12--19}.

\bibitem[Huang {\em et~al.\/}(2013{\natexlab{{\em a\/}}})Huang, Wu \&
  Cheng]{huang2013new}
{\sc \au{Huang, R.}, \au{Wu, H.} \& \au{Cheng, P.}} \yr{2013{\natexlab{{\em
  a\/}}}}  \at{A new lattice {B}oltzmann model for solid--liquid phase change}.
   \jt{Int. J. Heat Mass Transf.}  \bvol{59},  \pg{295--301}.

\bibitem[Huang {\em et~al.\/}(2013{\natexlab{{\em b\/}}})Huang, Kaczorowski,
  Ni, Xia {\em et~al.\/}]{huang2013confinement}
{\sc \au{Huang, S.-D.}, \au{Kaczorowski, M.}, \au{Ni, R.}, \au{Xia, K.-Q.} \&
  \au{others}} \yr{2013{\natexlab{{\em b\/}}}}  \at{Confinement-induced
  heat-transport enhancement in turbulent thermal convection}.  \jt{Phys. Rev.
  Lett.}  \bvol{111}~(10),  \pg{104501}.

\bibitem[Huang \& Xia(2016)]{huang2016effects}
{\sc \au{Huang, S.-D.} \& \au{Xia, K.-Q.}} \yr{2016}  \at{Effects of geometric
  confinement in quasi-2-{D} turbulent {R}ayleigh-{B}{\'e}nard convection}.
  \jt{J. Fluid Mech.}  \bvol{794},  \pg{639--654}.

\bibitem[Jebasingh \& Arasu(2020)]{jebasingh2020comprehensive}
{\sc \au{Jebasingh, B.~E.} \& \au{Arasu, A.~V.}} \yr{2020}  \at{A comprehensive
  review on latent heat and thermal conductivity of nanoparticle dispersed
  phase change material for low-temperature applications}.  \jt{Energy Storage
  Mater.}  \bvol{24},  \pg{52--74}.

\bibitem[Kean {\em et~al.\/}(2019)Kean, Sidik \& Kaur]{kean2019numerical}
{\sc \au{Kean, T.~H.}, \au{Sidik, N. A.~C.} \& \au{Kaur, J.}} \yr{2019}
  Numerical investigation on melting of phase change material ({PCM}) dispersed
  with various nanoparticles inside a square enclosure.  \bt{In {\em IOP Conf.
  Ser. Mater. Sci. Eng.\/}}, ,  \vol{vol. 469},  \pg{p. 012034}. IOP
  Publishing.

\bibitem[Khan \& Singh(2024)]{khan2024review}
{\sc \au{Khan, J.} \& \au{Singh, P.}} \yr{2024}  \at{Review on phase change
  materials for spacecraft avionics thermal management}.  \jt{J. Energy
  Storage}  \bvol{87},  \pg{111369}.

\bibitem[Khan \& Khan(2019)]{khan2019thermodynamic}
{\sc \au{Khan, Z.} \& \au{Khan, Z.~A.}} \yr{2019}  \at{Thermodynamic
  performance of a novel shell-and-tube heat exchanger incorporating paraffin
  as thermal storage solution for domestic and commercial applications}.
  \jt{Appl. Therm. Eng.}  \bvol{160},  \pg{114007}.

\bibitem[Kibria {\em et~al.\/}(2015)Kibria, Anisur, Mahfuz, Saidur \&
  Metselaar]{kibria2015review}
{\sc \au{Kibria, M.~A.}, \au{Anisur, M.~R.}, \au{Mahfuz, M.~H.}, \au{Saidur,
  R.} \& \au{Metselaar, I. H. S.~C.}} \yr{2015}  \at{A review on thermophysical
  properties of nanoparticle dispersed phase change materials}.  \jt{Energ.
  Convers. Manage.}  \bvol{95},  \pg{69--89}.

\bibitem[Kim {\em et~al.\/}(2008)Kim, Lee \& Choi]{kim2008onset}
{\sc \au{Kim, M.~C.}, \au{Lee, D.~W.} \& \au{Choi, C.~K.}} \yr{2008}  \at{Onset
  of buoyancy-driven convection in melting from below}.  \jt{Korean J. Chem.
  Eng.}  \bvol{25}~(6),  \pg{1239--1244}.

\bibitem[Kishore {\em et~al.\/}(2023)Kishore, Mahvi, Singh \&
  Woods]{kishore2023finned}
{\sc \au{Kishore, R.~A.}, \au{Mahvi, A.}, \au{Singh, A.} \& \au{Woods, J.}}
  \yr{2023}  \at{Finned-tube-integrated modular thermal storage systems for
  hvac load modulation in buildings}.  \jt{Cell Rep. Phys. Sci.}
  \bvol{4}~(12),  \pg{101704}.

\bibitem[Korti \& Guellil(2020)]{korti2020experimental}
{\sc \au{Korti, A. I.~N.} \& \au{Guellil, H.}} \yr{2020}  \at{Experimental
  study of the effect of inclination angle on the paraffin melting process in a
  square cavity}.  \jt{J. Energy Storage}  \bvol{32},  \pg{101726}.

\bibitem[Lachheb {\em et~al.\/}(2024)Lachheb, Younsi, Youssef \&
  Bouadila]{lachheb2024enhancing}
{\sc \au{Lachheb, M.}, \au{Younsi, Z.}, \au{Youssef, N.} \& \au{Bouadila, S.}}
  \yr{2024}  \at{Enhancing building energy efficiency and thermal performance
  with {PCM}-integrated brick walls: A comprehensive review}.  \jt{Build.
  Environ.}  \pg{p. 111476}.

\bibitem[Levin {\em et~al.\/}(2013)Levin, Shitzer \&
  Hetsroni]{levin2013numerical}
{\sc \au{Levin, P.~P.}, \au{Shitzer, A.} \& \au{Hetsroni, G.}} \yr{2013}
  \at{Numerical optimization of a {PCM}-based heat sink with internal fins}.
  \jt{Int. J. Heat Mass Transf.}  \bvol{61},  \pg{638--645}.

\bibitem[Li {\em et~al.\/}(2022)Li, Jiao \& Jia]{li2022melting}
{\sc \au{Li, M.}, \au{Jiao, Z.} \& \au{Jia, P.}} \yr{2022}  \at{Melting
  processes of phase change materials in a horizontally placed rectangular
  cavity}.  \jt{J. Fluid Mech.}  \bvol{950},  \pg{A34}.

\bibitem[Li \& Su(2023)]{li2023melting}
{\sc \au{Li, Y.} \& \au{Su, G.}} \yr{2023}  \at{Melting processes of phase
  change material in sidewall-heated cavity}.  \jt{J. Thermophys. Heat
  Transfer}  \bvol{37}~(2),  \pg{513--518}.

\bibitem[Liu {\em et~al.\/}(2022)Liu, Zheng \& Li]{liu2022high}
{\sc \au{Liu, Y.}, \au{Zheng, R.} \& \au{Li, J.}} \yr{2022}  \at{High latent
  heat phase change materials ({PCM}s) with low melting temperature for thermal
  management and storage of electronic devices and power batteries: Critical
  review}.  \jt{Renew. Sust. Energ.}  \bvol{168},  \pg{112783}.

\bibitem[Lu {\em et~al.\/}(2022)Lu, Zhang, Luo, Wu \& Yi]{lu2022rayleigh}
{\sc \au{Lu, C.}, \au{Zhang, M.}, \au{Luo, K.}, \au{Wu, J.} \& \au{Yi, H.}}
  \yr{2022}  \at{{R}ayleigh-{B}{\'e}nard instability in the presence of phase
  boundary and shear}.  \jt{J. Fluid Mech.}  \bvol{948},  \pg{A46}.

\bibitem[Luo {\em et~al.\/}(2015)Luo, Yao, Yi \& Tan]{luo2015lattice}
{\sc \au{Luo, K.}, \au{Yao, F.-J.}, \au{Yi, H.-L.} \& \au{Tan, H.-P.}}
  \yr{2015}  \at{Lattice {B}oltzmann simulation of convection melting in
  complex heat storage systems filled with phase change materials}.  \jt{Appl.
  Therm. Eng.}  \bvol{86},  \pg{238--250}.

\bibitem[Madruga \& Curbelo(2018)]{madruga2018dynamic}
{\sc \au{Madruga, S.} \& \au{Curbelo, J.}} \yr{2018}  \at{Dynamic of plumes and
  scaling during the melting of a phase change material heated from below}.
  \jt{Int. J. Heat Mass Transf.}  \bvol{126},  \pg{206--220}.

\bibitem[Purseed {\em et~al.\/}(2020)Purseed, Favier, Duchemin \&
  Hester]{purseed2020bistability}
{\sc \au{Purseed, J.}, \au{Favier, B.}, \au{Duchemin, L.} \& \au{Hester,
  E.~W.}} \yr{2020}  \at{Bistability in {R}ayleigh-{B}{\'e}nard convection with
  a melting boundary}.  \jt{Phys. Rev. Fluids}  \bvol{5}~(2),  \pg{023501}.

\bibitem[Ragoowansi {\em et~al.\/}(2023)Ragoowansi, Garimella \&
  Goyal]{ragoowansi2023realistic}
{\sc \au{Ragoowansi, E.~A.}, \au{Garimella, S.} \& \au{Goyal, A.}} \yr{2023}
  \at{Realistic utilization of emerging thermal energy recovery and storage
  technologies for buildings}.  \jt{Cell Rep. Phys. Sci.}  \bvol{4}~(5),
  \pg{101393}.

\bibitem[Regin {\em et~al.\/}(2009)Regin, Solanki \& Saini]{regin2009analysis}
{\sc \au{Regin, A.~F.}, \au{Solanki, S.~C.} \& \au{Saini, J.~S.}} \yr{2009}
  \at{An analysis of a packed bed latent heat thermal energy storage system
  using {PCM} capsules: Numerical investigation}.  \jt{Renew. Energy}
  \bvol{34}~(7),  \pg{1765--1773}.

\bibitem[Shishkina(2016)]{shishkina2016momentum}
{\sc \au{Shishkina, O.}} \yr{2016}  \at{Momentum and heat transport scalings in
  laminar vertical convection}.  \jt{Phys. Rev. E}  \bvol{93}~(5),
  \pg{051102}.

\bibitem[Shishkina(2021)]{shishkina2021rayleigh}
{\sc \au{Shishkina, O.}} \yr{2021}  \at{Rayleigh-{B}{\'e}nard convection: The
  container shape matters}.  \jt{Phys. Rev. Fluids}  \bvol{6}~(9),
  \pg{090502}.

\bibitem[Sun {\em et~al.\/}(2016)Sun, Zhang, Medina \&
  Lee]{sun2016experimental}
{\sc \au{Sun, X.}, \au{Zhang, Q.}, \au{Medina, M.~A.} \& \au{Lee, K.~O.}}
  \yr{2016}  \at{Experimental observations on the heat transfer enhancement
  caused by natural convection during melting of solid--liquid phase change
  materials ({PCM}s)}.  \jt{Appl. Energy}  \bvol{162},  \pg{1453--1461}.

\bibitem[Tariq {\em et~al.\/}(2020)Tariq, Ali, Akram, Janjua \&
  Ahmadlouydarab]{tariq2020nanoparticles}
{\sc \au{Tariq, S.~L.}, \au{Ali, H.~M.}, \au{Akram, M.~A.}, \au{Janjua, M.~M.}
  \& \au{Ahmadlouydarab, M.}} \yr{2020}  \at{Nanoparticles enhanced phase
  change materials ({N}e{PCM}s)-{A} recent review}.  \jt{Appl. Therm. Eng.}
  \bvol{176},  \pg{115305}.

\bibitem[Tong {\em et~al.\/}(2021)Tong, Nie, Li, Li, Zou, Jiang, Jin \&
  Ding]{tong2021phase}
{\sc \au{Tong, S.}, \au{Nie, B.}, \au{Li, Z.}, \au{Li, C.}, \au{Zou, B.},
  \au{Jiang, L.}, \au{Jin, Y.} \& \au{Ding, Y.}} \yr{2021}  \at{A phase change
  material ({PCM}) based passively cooled container for integrated road-rail
  cold chain transportation---an experimental study}.  \jt{Appl. Therm. Eng.}
  \bvol{195},  \pg{117204}.

\bibitem[Toppaladoddi \& Wettlaufer(2019)]{toppaladoddi2019combined}
{\sc \au{Toppaladoddi, S.} \& \au{Wettlaufer, J.~S.}} \yr{2019}  \at{The
  combined effects of shear and buoyancy on phase boundary stability}.  \jt{J.
  Fluid Mech.}  \bvol{868},  \pg{648--665}.

\bibitem[Vajjha {\em et~al.\/}(2010)Vajjha, Das \&
  Namburu]{vajjha2010numerical}
{\sc \au{Vajjha, R.~S.}, \au{Das, D.~K.} \& \au{Namburu, P.~K.}} \yr{2010}
  \at{Numerical study of fluid dynamic and heat transfer performance of
  {A}l$_2${O}$_3$ and {C}u{O} nanofluids in the flat tubes of a radiator}.
  \jt{Int. J. Heat Fluid Fl.}  \bvol{31}~(4),  \pg{613--621}.

\bibitem[Vasil \& Proctor(2011)]{vasil2011dynamic}
{\sc \au{Vasil, G.~M.} \& \au{Proctor, M. R.~E.}} \yr{2011}  \at{Dynamic
  bifurcations and pattern formation in melting-boundary convection}.  \jt{J.
  Fluid Mech.}  \bvol{686},  \pg{77--108}.

\bibitem[Verma {\em et~al.\/}(2008)Verma, Singal {\em
  et~al.\/}]{verma2008review}
{\sc \au{Verma, P.}, \au{Singal, S.~K.} \& \au{others}} \yr{2008}  \at{Review
  of mathematical modeling on latent heat thermal energy storage systems using
  phase-change material}.  \jt{Renew. Sustain. Energy Rev.}  \bvol{12}~(4),
  \pg{999--1031}.

\bibitem[Vogel {\em et~al.\/}(2016)Vogel, Felbinger \&
  Johnson]{vogel2016natural}
{\sc \au{Vogel, J.}, \au{Felbinger, J.} \& \au{Johnson, M.}} \yr{2016}
  \at{Natural convection in high temperature flat plate latent heat thermal
  energy storage systems}.  \jt{Appl. Energy}  \bvol{184},  \pg{184--196}.

\bibitem[Voller {\em et~al.\/}(1987)Voller, Cross \&
  Markatos]{voller1987enthalpy}
{\sc \au{Voller, V.~R.}, \au{Cross, M.} \& \au{Markatos, N.~C.}} \yr{1987}
  \at{An enthalpy method for convection/diffusion phase change}.  \jt{Int. J.
  Numer. Methods Biomed. Eng.}  \bvol{24}~(1),  \pg{271--284}.

\bibitem[Wang {\em et~al.\/}(2012)Wang, Ma, Chen \& Sun]{wang2012linear}
{\sc \au{Wang, B.-F.}, \au{Ma, D.-J.}, \au{Chen, C.} \& \au{Sun, D.-J.}}
  \yr{2012}  \at{Linear stability analysis of cylindrical
  {R}ayleigh-{B}{\'e}nard convection}.  \jt{J. Fluid Mech.}  \bvol{711},
  \pg{27--39}.

\bibitem[Wang {\em et~al.\/}(2022{\natexlab{{\em a\/}}})Wang, Peng, Peng \&
  Zhang]{wang2022fluidic}
{\sc \au{Wang, H.}, \au{Peng, Y.}, \au{Peng, H.} \& \au{Zhang, J.}}
  \yr{2022{\natexlab{{\em a\/}}}}  \at{Fluidic phase--change materials with
  continuous latent heat from theoretically tunable ternary metals for
  efficient thermal management}.  \jt{Proc. Natl. Acad. Sci. U.S.A.}
  \bvol{119}~(31),  \pg{e2200223119}.

\bibitem[Wang {\em et~al.\/}(2021)Wang, Liu, Verzicco, Shishkina \&
  Lohse]{wang2021regime}
{\sc \au{Wang, Q.}, \au{Liu, H.-R.}, \au{Verzicco, R.}, \au{Shishkina, O.} \&
  \au{Lohse, D.}} \yr{2021}  \at{Regime transitions in thermally driven
  high-{R}ayleigh number vertical convection}.  \jt{J. Fluid Mech.}
  \bvol{917},  \pg{A6}.

\bibitem[Wang {\em et~al.\/}(2022{\natexlab{{\em b\/}}})Wang, Li, Luo, Wang \&
  Shah]{wang2022critical}
{\sc \au{Wang, X.}, \au{Li, W.}, \au{Luo, Z.}, \au{Wang, K.} \& \au{Shah,
  S.~P.}} \yr{2022{\natexlab{{\em b\/}}}}  \at{A critical review on phase
  change materials ({PCM}) for sustainable and energy efficient building:
  Design, characteristic, performance and application}.  \jt{Energy Build.}
  \bvol{260},  \pg{111923}.

\bibitem[Wang {\em et~al.\/}(1999)Wang, Amiri \& Vafai]{wang1999experimental}
{\sc \au{Wang, Y}, \au{Amiri, A} \& \au{Vafai, K}} \yr{1999}  \at{An
  experimental investigation of the melting process in a rectangular
  enclosure}.  \jt{Int. J. Heat Mass Transf.}  \bvol{42}~(19),
  \pg{3659--3672}.

\bibitem[Weideman \& Reddy(2000)]{weideman2000matlab}
{\sc \au{Weideman, J~Andre} \& \au{Reddy, Satish~C}} \yr{2000}  \at{A {MATLAB}
  differentiation matrix suite}.  \jt{ACM Trans. Math. Softw.}  \bvol{26}~(4),
  \pg{465--519}.

\bibitem[Woods {\em et~al.\/}(2021)Woods, Mahvi, Goyal, Kozubal, Odukomaiya \&
  Jackson]{woods2021rate}
{\sc \au{Woods, J.}, \au{Mahvi, A.}, \au{Goyal, A.}, \au{Kozubal, E.},
  \au{Odukomaiya, A.} \& \au{Jackson, R.}} \yr{2021}  \at{Rate capability and
  {R}agone plots for phase change thermal energy storage}.  \jt{Nat. Energy}
  \bvol{6}~(3),  \pg{295--302}.

\bibitem[Xia {\em et~al.\/}(2024)Xia, Fu, Feng, Gong \& Yu]{xia2024particle}
{\sc \au{Xia, M.}, \au{Fu, J.}, \au{Feng, Y.~T.}, \au{Gong, F.} \& \au{Yu, J.}}
  \yr{2024}  \at{A particle-resolved heat-particle-fluid coupling model by
  {DEM}-{IMB}-{LBM}}.  \jt{Int. J. Rock Mech. Min. Sci.}  \bvol{16}~(6),
  \pg{2267--2281}.

\bibitem[Yang {\em et~al.\/}(2020)Yang, Huang \& Zhou]{yang2020thermophysical}
{\sc \au{Yang, L.}, \au{Huang, J.-N.} \& \au{Zhou, F.}} \yr{2020}
  \at{Thermophysical properties and applications of nano-enhanced {PCM}s: An
  update review}.  \jt{Energ. Convers. Manage.}  \bvol{214},  \pg{112876}.

\bibitem[Yang {\em et~al.\/}(2022)Yang, Chong, Liu, Verzicco \&
  Lohse]{yang2022abrupt}
{\sc \au{Yang, R.}, \au{Chong, K.~L.}, \au{Liu, H.-R.}, \au{Verzicco, R.} \&
  \au{Lohse, D.}} \yr{2022}  \at{Abrupt transition from slow to fast melting of
  ice}.  \jt{Phys. Rev. Fluids}  \bvol{7}~(8),  \pg{083503}.

\bibitem[Yang {\em et~al.\/}(2021)Yang, King \& Miljkovic]{yang2021phase}
{\sc \au{Yang, T.}, \au{King, W.~P.} \& \au{Miljkovic, N.}} \yr{2021}
  \at{Phase change material-based thermal energy storage}.  \jt{Cell Rep. Phys.
  Sci.}  \bvol{2}~(8),  \pg{100540}.

\bibitem[Yu {\em et~al.\/}(2017)Yu, Goldfaden, Flagstad \& Scheel]{yu2017onset}
{\sc \au{Yu, J.}, \au{Goldfaden, A.}, \au{Flagstad, M.} \& \au{Scheel, J.~D.}}
  \yr{2017}  \at{Onset of {R}ayleigh-{B}{\'e}nard convection for intermediate
  aspect ratio cylindrical containers}.  \jt{Phys. Fluids}  \bvol{29}~(2),
  \pg{024107}.

\bibitem[Zhang {\em et~al.\/}(2023)Zhang, Tang, Chen, Zhang, Chen, Ding, Zhou,
  Xu, Liu \& Xue]{zhang2023accelerating}
{\sc \au{Zhang, Y.}, \au{Tang, J.}, \au{Chen, J.}, \au{Zhang, Y.}, \au{Chen,
  X.}, \au{Ding, M.}, \au{Zhou, W.}, \au{Xu, X.}, \au{Liu, H.} \& \au{Xue, G.}}
  \yr{2023}  \at{Accelerating the solar-thermal energy storage via inner-light
  supplying with optical waveguide}.  \jt{Nat. Commun.}  \bvol{14}~(1),
  \pg{3456}.

\bibitem[Zhou {\em et~al.\/}(2023)Zhou, Xu \& Huang]{zhou2023adaptive}
{\sc \au{Zhou, X.}, \au{Xu, X.} \& \au{Huang, J.}} \yr{2023}  \at{Adaptive
  multi-temperature control for transport and storage containers enabled by
  phase-change materials}.  \jt{Nat. Commun.}  \bvol{14}~(1),  \pg{5449}.

\end{thebibliography}



\end{document}